\documentclass{article}

\usepackage{arxiv}

\usepackage[utf8]{inputenc} 
\usepackage[T1]{fontenc}    
\usepackage{hyperref}       
\usepackage{url}            
\usepackage{booktabs}       
\usepackage{amsfonts}       
\usepackage{nicefrac}       
\usepackage{microtype}      
\usepackage{lipsum}
\usepackage{enumitem}
\usepackage{graphicx}
\usepackage{listings}
\usepackage{multirow}
\usepackage{makecell} 
\usepackage{bbding}
\usepackage{subfigure}
\usepackage{caption}
\usepackage[ruled,vlined]{algorithm2e}
\usepackage{tabularx}
\usepackage[table]{xcolor}
\definecolor{ForestGreen}{HTML}{006400}
\definecolor{Maroon}{HTML}{8B0000}

\renewcommand{\paragraph}[1]{\vskip 0.05in \noindent\textbf{#1.}}
\newcommand{\tool}{DTVM\xspace}

\setlist[itemize]{itemindent=0.01em}

\hypersetup{hidelinks}
\graphicspath{ {./images/} }

\title{DTVM: Revolutionizing Smart Contract Execution with Determinism and Compatibility \\ \small{Featuring EVM Compatibility, Multi-Language Support, Diverse Security Hardware Integration, and AI-Ready Extensions for Enhanced Smart Contract Execution}}

\author{
\begin{minipage}[t]{\textwidth}
\raggedright
\normalfont Wei Zhou, Xiong Xu, Changzheng Wei,  Ying Yan, Wei Tang, Zhihao Chen, Xuebing Huang, Wengang Chen, 
\\Jie Zhang, Yang Chen, Xiaofu Zheng, Hanghang Wu, Shenglong Chen, Ermei Wang, Xiangfei Chen, Yang Yu, 
\\Meng Wu, Tao Zhu, Liwei Yuan, Feng Yu, Alex Zhang, Wei Wang, Ji Luo, Zhengyu He, and Wenbiao Zhao
\end{minipage}
}

\usepackage{cite}
\begin{document}
\maketitle
\begin{abstract}
We introduce the \textbf{D}e\textbf{T}erministic \textbf{V}irtual \textbf{M}achine (DTVM) Stack, a next-generation smart contract execution framework designed to address critical performance, determinism, and ecosystem compatibility challenges in blockchain networks. Building upon WebAssembly (Wasm) while maintaining full Ethereum Virtual Machine (EVM) ABI compatibility, DTVM introduces a \textbf{D}eterministic \textbf{M}iddle \textbf{I}ntermediate \textbf{R}epresentation (dMIR) and a hybrid lazy-JIT compilation engine that dynamically adapts optimization levels (O0$\sim$O2) to balance compilation speed and execution efficiency. 
Beyond Wasm, DTVM further accommodates diverse instruction set architectures (e.g., EVM, RISC-V) through modular adaptation layers, translating their bytecode into the unified dMIR intermediate representation. This enables seamless integration with DTVM’s hybrid lazy-JIT compilation engine, which dynamically optimizes performance while preserving deterministic execution guarantees across heterogeneous environments.
The \textit{key contributions} including:
\textbf{1)}. The framework achieves  up to 2$\times$ acceleration over evmone in dominant Ethereum contract (e.g. ERC20/721/1155) execution and reduces fibonacci computation latency by 11.8$\sim$40.5\% compared to Wasm based VMs, while ensuring cross-architecture deterministic execution through hardware-enforced memory isolation and standardized trap handling. 
\textbf{2)}. A novel trampoline hot-switch mechanism enables sub-millisecond (0.95ms) post-deployment invocation times, outperforming up to about 23$\times$ in compilation and invocation efficiency comparing to the state-of-the-art Wasm  (Wasmtime).
\textbf{3)}. The solution supports multi-language development (Solidity, C++, Rust, Java, Go, and AssemblyScript) through unified bytecode conversion while maintaining EVM ABI compatibility for seamless invocation. 
Additionally, it reduces machine code object sizes by 30.0$\sim$72.6\% compared to alternatives, coupled with a minimized Trusted Computing Base (69.5KLoC, 48\% of Wasmtime) for secure execution environments. 
\textbf{4)}. DTVM Stack offers \textbf{SmartCogent}, an AI-driven full-stack development experience, leveraging fine-tuned LLMs and retrieval-augmented generation to automate tasks across the smart contract lifecycle: development, debugging, security auditing, and deployment. Security auditing of SmartCogent archives 81\% vulnerability detection accuracy and 86\% automated repair success rates. Experimental validation across PolyBench (a general Wasm testbench) benchmarks and Ethereum ecosystem workloads demonstrates DTVM’s capability to deliver deterministic, high-throughput execution while maintaining backward compatibility with EVM-based systems, positioning it as a foundational infrastructure for scalable Web3 applications. 
The DTVM Stack has been open-sourced\footnote{https://github.com/DTVMStack}, comprising its core execution engine, multi-language SDKs, and associated toolchain components.
\end{abstract}


\section{Introduction}
\label{sec:intro}
Over the past decade, the infrastructure underpinning blockchain systems has undergone remarkable advancements, giving rise to a vibrant ecosystem of decentralized applications. These innovations  have driven the emergence of more robust, scalable, and secure platforms, empowering decentralized applications to thrive within a rapidly evolving ecosystem. This rapid evolution has brought about significant performance and scalability challenges, prompting continuous innovation in high-performance blockchain systems$-$a trend that has become increasingly pronounced in recent years. Central to these challenges are the inherent bottlenecks on smart contract execution, highlighting the necessity for high-performance smart contract virtual machines (VMs) that are essential for the future of Web3.

In blockchain systems, ensuring determinism in distributed execution is crucial. Traditionally, stack-based VMs relying on interpretation have been the preferred approach, with Ethereum’s EVM (Ethereum Virtual Machine)~\cite{wood2014ethereum} serving as the pivotal role. However, as performance demands escalate dramatically, there is a growing need for enhancements that build upon and extend the core capabilities of the EVM. Several high-performance blockchain systems propose various solutions from the perspectives of introducing new smart contract virtual machines~\cite{Sui,Aptos,blocks2021introduction}, optimizing storage access~\cite{tian2024letus}, and developing new parallel execution architectures~\cite{Pharos,Monad,MegaETH}, to enhance the overall system throughput. 

In this work, we introduce \textbf{D}e\textbf{T}erministic \textbf{V}irtual \textbf{M}achine (DTVM) Stack, which places a comprehensive solution for smart contract execution. The DTVM Stack has been open-sourced, featuring a high-performance execution engine designed to enhance smart contract execution efficiency. This stack introduces a function-level lazy Just-In-Time (JIT) compilation framework specifically tailored for smart contracts. It ensures full compatibility with the EVM ABI while supporting various contract development languages, including C/C++, Rust, Java, Go, and AssemblyScript, in addition to Solidity. Furthermore, the DTVM Stack provides native portability for Trusted Execution Environments (TEEs) and integrates a high-precision AI-driven auditing agent. It also offers an intelligent development framework that covers the entire lifecycle of smart contract creation, debugging, and deployment, thereby streamlining the process from initial code generation to final deployment. \tool Stack enables seamless integration with the broader EVM ecosystem, allowing developers to easily migrate their applications with minimal effort. Through this work, We aim to contribute to the EVM community by sharing insights and optimizations that can practically help improve the overall performance and efficiency of EVM-based blockchain systems.

Building upon its foundational support for WebAssembly(Wasm) and EVM ABI compatibility, DTVM further demonstrates architectural extensibility through its modular design. Beyond Wasm, DTVM could extend its architecture to support diverse instruction set architectures (e.g., EVM, RISC-V) through modular adaptation layers. Notably, RISC-V has emerged as a prominent candidate instruction set for zero-knowledge virtual machines (ZKVMs)~\cite{risc0, sp1, valida,arun2024jolt}, with widespread adoption in community implementations. Discussions within the Ethereum community are actively exploring its potential~\cite{risv-vm} as the underlying architecture for next-generation smart contract virtual machines—a direction closely aligned with DTVM’s forward-looking design, which provides a high-performance execution engine for diverse instruction set architectures and supports future integration of zero-knowledge (ZK) capabilities.

This paper delves into our next-generation smart contract VM, which systematically addresses and overcomes the execution bottlenecks in current blockchain systems. By extending the strengths of the EVM, \tool promises to deliver unprecedented performance and scalability, paving the way for a more efficient and scalable decentralized future. 

The key contributions are outlined as follows:
\begin{itemize}
    \item[1.] \textbf{Deterministic JIT Execution Engine with Enhanced Performance}: Central to the DTVM Stack is an ultra-fast execution engine that employs deterministic JIT compilation. This engine implements a unique function-level lazy JIT compilation framework, which enables asynchronous compilation at the granularity of individual functions. To address the inherent trade-off between compilation time and execution performance in traditional JIT processes, \tool introduces a hybrid compilation strategy, which dynamically adapts and switches between O0 and O2 compilation optimization levels during runtime, thereby optimizing both compilation efficiency and execution speed. Additionally, the DTVM Stack proposes Deterministic Middle Intermediate Representation (dMIR), a blockchain-specific intermediate representation that ensures deterministic execution guarantees. By carefully analyzing non-deterministic behaviors in Wasm runtime, \tool identifies three major categories of non-determinism: \textit{Static and loading-time non-determinism}, \textit{Runtime non-determinism} and \textit{Exception and trap non-determinism}. To eliminate these non-determinism, \tool propose several novel mechanisms: \textit{Deterministic Halt Mechanism}, \textit{Numerical Computation Determinism }, \textit{Deterministic Error Handling}, \textit{Deterministic Format Validation} and \textit{Stack Determinism}. These mechanisms collectively ensure predictable and strict deterministic execution outcomes.
    \item[2.] \textbf{EVM ABI Compatibility and Multi-Language Ecosystem Support}: The DTVM Stack maintains compatibility with the latest Solidity 0.8.x specification while expanding support to six frontend programming languages (e.g. Solidity, C/C++, Rust, Java, Golang, and AssemblyScript). The unified contract bytecode intermediate representation enables efficient conversion between EVM bytecode and other language-specific formats. Benefiting from these designs, DTVM Stack facilitates cross-language interoperability, enabling seamless coordination between different contract implementations and enhancing the flexibility of smart contract engineering.
    \item[3.] \textbf{TEE-Native Security and Hardware-Optimized Efficiency}: For application-level TEEs such as Intel SGX~\cite{costan2016intel}, the DTVM Stack offers high portability through a minimized Trusted Computing Base (TCB)~\cite{yan2020confidentiality}. Compared to competitive Wasm implementations, the DTVM Stack reduces codebase size to 48\% (comparing to Wasmtime) and binary library size to approximately 60\%, thereby minimizing potential attack surfaces while maintaining security and efficiency. Additionally, the framework leverages modern processor registers and exception handling mechanisms to address specialized requirements such as gas metering and boundary checks in JIT compilation. This design effectively balances security and performance by ensuring efficient resource utilization without compromising execution safety.
    \item[4.] \textbf{ AI-Powered Smart Contract Development and Auditing}: SmartCogent is developed based on a multi-agent architecture. It automates critical stages of the smart contract lifecycle, including code generation, compilation, testing, deployment, and security auditing. Leveraging advanced Retrieval-Augmented Generation (RAG) data and Large Language Model (LLM) capabilities, it enhances the developer experience by introducing intelligent automation into the development process.   
\end{itemize}

Significant advantages are demonstrated by DTVM in deterministic smart contract execution efficiency, post-deployment invocation latency, multi-language ecosystem support, and intelligent development through comprehensive benchmarking. \textbf{1). Superior Deterministic Execution Efficiency:} Performance improvements of up to 2$\times$ are achieved in dominant Ethereum ERCs (e.g. ERC20/721/1155) contract execution compared to evmone, with a 3.61$\times$ acceleration observed for compute-intensive Fibonacci workloads. The JIT compilation mode further delivers a 58.54$\times$ performance gain over the interpreter baseline. When evaluated against mainstream Wasm VMs, the framework exhibits an average speedup ratio of 1.20$\sim$12.14$\times$ across the PolyBench standard bench suite, while reducing Fibonacci processing time by 11.8$\sim$40.5\%, with cross-architecture deterministic execution guarantee in all tested scenarios. \textbf{2). Sub-Millisecond Post-Deployment Invocation Latency: }The on-demand compilation mechanism reduces the post-deployment invocation time to 0.95ms in PolyBench cases, achieving over 23$\times$ speedup compared to Wasmtime (Cranelift as JIT backend). This rapid invocation capability is critical for blockchain transaction processing, as prolonged loading times can induce notable performance degradation and expose exploitable vulnerabilities for DDoS attacks. \textbf{3). Enhanced Functionality with Full Compatibility:} In code generation metrics, machine code object sizes are reduced by 30$\sim$72\% relative to comparable solutions, with the core codebase (69.5KLoC) representing only 48\% of Wasmtime’s implementation. The CLI binary footprint (32MB) is minimized by up to 71\% compared to baseline systems, establishing a minimized Trusted Computing Base (TCB) for secure execution environments. Functional validation confirms DTVM’s dual capability of maintaining full Ethereum ecosystem compatibility while establishing a multi-language smart contract development framework supporting C/C++, Rust, Java, Go, and AssemblyScript. \textbf{4). Intelligent Coding and High-precision Security Auditing: } \tool's integrated SmartCogent delivers an 81\% accuracy rate in vulnerability detection and an 86\% success rate in automated repair, significantly surpassing other baseline models and specialized tools. These extensive experimental results highlight the \tool Stack as an efficient and secure infrastructure execution framework for high-performance blockchain applications, providing optimized computational throughput with deterministic execution guarantees.

\section{Related Work}
\label{sec:related_work}
\subsection{WebAssembly}
\label{sec:webassembly}
WebAssembly (Wasm)~\cite{WebAssembly}, as a platform-independent binary instruction format, 
has become a pivotal technology in the blockchain smart contract domain due to its strong isolation, multi-language support, 
and efficient execution characteristics. Wasm VMs offer several advantages in blockchain environments:

\textbf{Multi-Language Ecosystem Support}: Wasm enables developers to write smart contracts in mainstream programming languages such as C/C++, Rust, Go, and Java, 
significantly lowering the barrier to dApp development. 
This flexibility allows existing developers to leverage familiar tools and languages in the Web3 space, fostering a broader ecosystem.

\textbf{High Execution Performance}: Compared to traditional interpreted VMs, 
Wasm is designed to achieve relative high execution speeds through its compact binary format and optimized execution model. This makes Wasm VMs particularly advantageous for handling complex computations.

\textbf{Portability and Security}: Wasm provides robust memory isolation and type safety guarantees while maintaining cross-platform portability. 
These features align well with blockchain requirements, ensuring consistent execution across different nodes.

Despite these advantages, applying Wasm VMs in blockchain environments presents unique challenges:

\textbf{Deterministic Execution Requirements}: Blockchain consensus mechanisms demand identical results across all nodes executing the same contract. However, the standard Wasm specification contains non-deterministic elements, such as floating-point operations, multithreading, and platform-dependent behaviors of certain instructions. 
  This necessitates strict controls and restrictions in blockchain-specific Wasm VM implementations.

\textbf{Resource Constraints and Gas Metering}: Blockchain environments require precise measurement and limitation of resource consumption (e.g., CPU time, memory usage) to prevent DoS attacks and resource abuse. This requires Wasm VMs to integrate gas metering mechanisms, assigning appropriate costs to each instruction while maintaining execution efficiency.

Currently, multiple blockchain projects utilize Wasm as their smart contract execution environment, including Near Protocol~\cite{Near}, Polkadot~\cite{Polkadot}, Cosmos (via CosmWasm)~\cite{CosmWasm}, and Arbitrum~\cite{Arbitrum}. 
Mainstream Wasm VMs include Wasmtime~\cite{Wasmtime} and Wasmer~\cite{Wasmer}. 
However, these VMs do not natively support the deterministic requirements of blockchain scenarios.
\subsection{JIT (Just-In-Time) Technology}
\label{sec:JIT}
Wasm code execution can be achieved through three primary methods: interpretation, JIT compilation~\cite{JIT}, and Ahead-of-Time (AoT) compilation~\cite{AOT}. 
Each method has distinct advantages and disadvantages in blockchain environments:

\textbf{Interpretation}: Offers fast startup times but suffers from low execution efficiency, making it unsuitable for computationally intensive contracts.

\textbf{JIT Compilation}: Balances startup speed with execution efficiency but consumes resources during the compilation process.

\textbf{AoT Compilation}: Provides the highest execution efficiency but increases contract deployment costs and on-chain storage burdens.

Notable projects include Wasmtime's Cranelift~\cite{eddine2024case} and Wasmer's integration of LLVM backends to provide JIT engine capabilities. 
In blockchain settings, JIT compilation has become the predominant choice due to its balanced characteristics. 
However, beyond security and determinism, achieving optimal trade-offs between compilation time and execution efficiency remains a challenge. 
Performing full JIT compilation during contract deployment can delay transactions, impacting user experience and network throughput. 
Particularly for complex contracts, compilation times may reach several seconds or longer. Additionally, JIT compilation faces the challenge of deterministic execution, requiring identical results across nodes, operating systems, and CPUs, even for failed transactions.

\subsection{EVM with JIT/AoT}
\label{sec:EVM_with_JIT}
Developers and researchers have actively explored methods to enhance EVM execution performance, including the adoption of Ahead-of-Time (AoT) and Just-In-Time (JIT) compilation techniques. A notable example is the revmc~\cite{revmc} virtual machine developed by the paradigm.xyz community, which converts smart contract code into mature intermediate representations. By leveraging the optimization capabilities of industrial-grade compiler frameworks, this approach generates highly efficient machine code. Specifically, EVM bytecode is first compiled into LLVM Intermediate Representation (LLVM IR), which is then executed using the LLVM JIT engine. This workflow significantly improves the performance of computation-intensive smart contracts by capitalizing on LLVM’s advanced optimization pipelines.

Furthermore, the evmone~\cite{evmone} interpreter employs a JIT-inspired "hot code caching" strategy to optimize execution. For frequently executed (hot) bytecode segments, evmone transforms them into contiguous instruction objects. This eliminates the need for repetitive loop-based interpretation—where execution cycles through opcode types and jumps between instructions—and instead processes cached instruction objects sequentially. This streamlined approach reduces overhead and achieves performance gains comparable to JIT compilation, while maintaining the simplicity of interpreter-based execution. Compared to traditional interpreter implementations, evmone's hybrid strategy demonstrates marked improvements in execution efficiency.

\subsection{LLM-Based Smart Contract Tools}
\label{sec:llm_based_smart_contract_tools}
Large language models (LLMs) have demonstrated significant advantages in code generation, 
such as efficient generation capabilities and keen contextual understanding, 
making them a focal point in academic research and practical applications~\cite{Zan,AntChainOpenLabs,AIforWeb3,PPL+22,JWS+24}. 
Smart contracts, as a core component of Web3, must balance security and efficiency. 
However, developing secure, reliable, and efficient smart contract code is a complex task. 
Recent studies have explored the use of LLMs for generating, debugging, and auditing smart contracts. 
For instance, some works have extended LLM-based code generation to Solidity smart contracts~\cite{CR25}. 
Others have fine-tuned LLMs to automatically generate comments for smart contracts, aiding developers in understanding contracts and reducing vulnerability risks~\cite{ZCYS24}. 
Research has also fine-tuned LLMs for automated vulnerability scanning of smart contracts, with some approaches generating repair solutions for defective code~\cite{MPM24,LXA+24}.

Compared to direct LLM applications, AI Agents offer real-time decision-making based on task states, providing automation and modularity advantages. 
For example, blockchain platforms like Solana employ AI Agents such as Solana Agent Kit and Eliza to automate tasks like transactions and token issuance~\cite{Solana+AI}. 
Near's AI Agent assists users in completing tasks like token swaps in the Web3 domain~\cite{Near+AI}. Blockchain analytics platforms like Dune~\cite{Dun25} have integrated AI tools to simplify SQL queries on blockchain data, enabling users to generate complex SQL statements from natural language inputs.

Using AI Agents for smart contract code generating, debugging, and auditing can significantly reduce development complexity and improve efficiency. 
Furthermore, combining AI Agents with Retrieval-Augmented Generation (RAG) ~\cite{LPP+20} allows dynamic referencing of external knowledge bases, enriching code generation with extensive knowledge support and further enhancing quality and precision.

\section{Architecture}
\label{sec:architecture}
DTVM is a full-stack development and execution engine tailored for the smart contract domain, offering deterministic execution guarantees while supporting Ethereum contract ecosystems. 
Its core features include multi-mode compilation and execution capabilities, high-performance engines, multi-language development support, and AI-augmented intelligent contract development, testing, and security auditing tools.

\begin{figure}[h!] 
  \centering
  \includegraphics[scale=0.5]{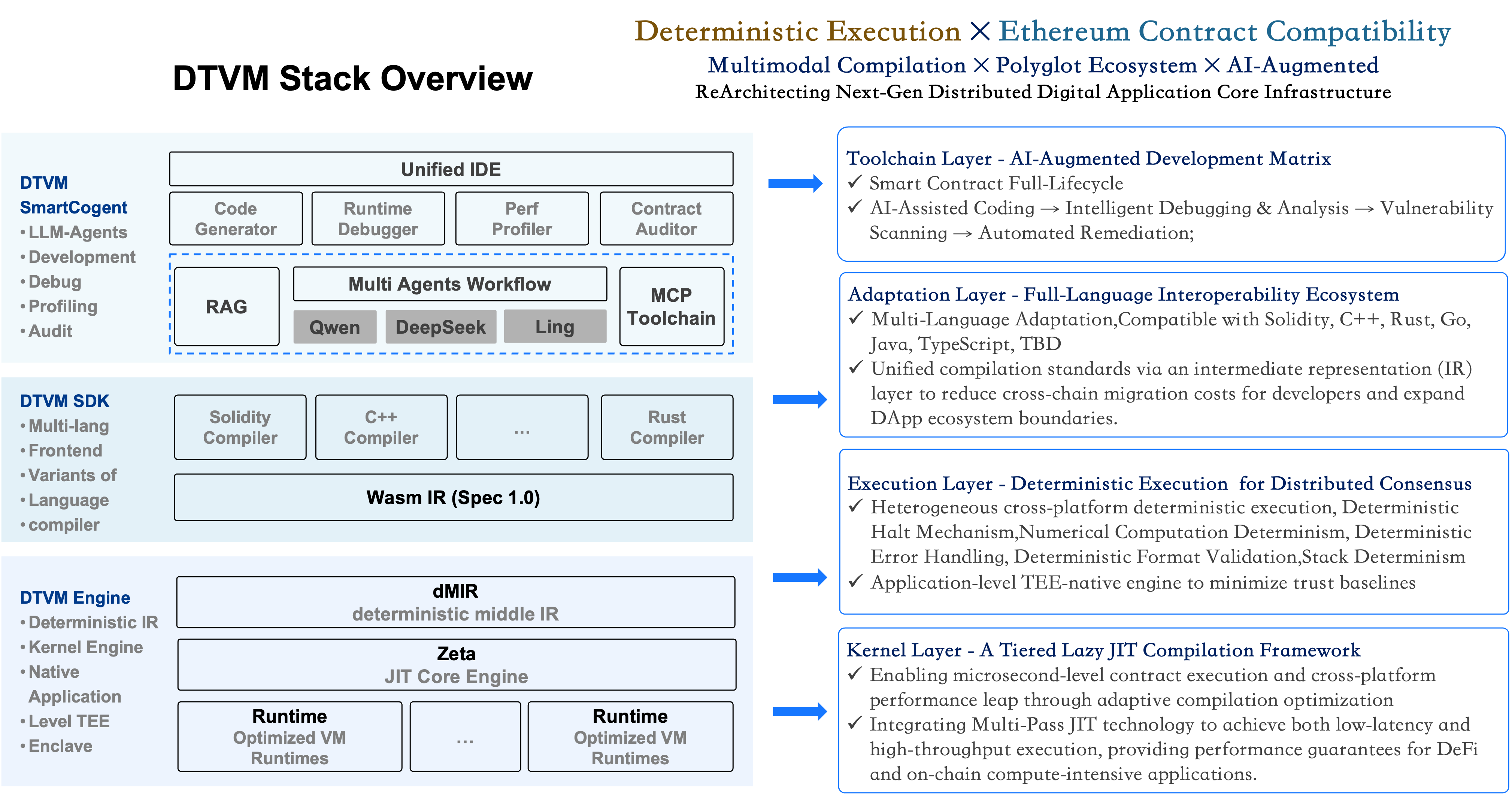}
  \caption{DTVM Stack Overview}
\end{figure}

The DTVM Stack's technical architecture is composed of three primary components: the DTVM Engine, DTVM SDK, and DTVM Tools. 
Logically, it is organized into four layers: the core layer, execution layer, adaptation layer, and toolchain layer. 
The core and execution layers are encapsulated within the DTVM Engine.
\subsection{DTVM Engine}
\label{sec:dtvm_engine}
The DTVM Engine introduces a Lazy-JIT compilation architecture and runtime environment for smart contract virtual machines, ensuring deterministic execution and high-performance capabilities. Internally, it comprises three core components: the dMIR, Zeta Engine, and VM Runtime.

\textbf{dMIR (Deterministic Middle IR)}: We propose a standard (T/TBI57-2024) which is approved in 2024, by defining and enforcing the deterministic WebAssembly (dWasm) extension specification. 
A deterministic intermediate representation (dMIR) is introduced. It unifies the abstraction layers of diverse smart contract VM intermediate representations (e.g., Wasm, EVM). 
It enforces deterministic execution mechanisms at the dMIR abstraction level, including: 1) Deterministic Halt Mechanism, 
2) Numerical Computation Determinism, 
3) Deterministic Error Handling, 
4) Deterministic Format Validation, 
5) Stack Determinism. 

This layer enables cross-platform deterministic execution and provides abstraction-level optimizations for contract logic and platform-agnostic code. 
Its plug-in architecture supports scalable extensions, while annotations and guidance are embedded to facilitate platform-specific optimizations in subsequent stages.

\textbf{Zeta Engine}: Traditional JIT execution prioritizes post-optimization execution efficiency, but blockchain environments necessitate balancing compilation time and performance. 
Single-pass compilation achieves rapid execution (O0 optimization level) but limits efficiency gains, while multi-pass strategies (O2 levels) enhance performance but introduce excessive compilation delays. 
The Zeta Engine addresses this trade-off through three core mechanisms:
\begin{itemize}[itemindent=0.1em]
  \item \textbf{Function-Granularity Compilation Partitioning}: Splits compilation tasks by function granularity instead of full-contract compilation, minimizing individual task loads.
  \item \textbf{Asynchronous Parallel Compilation}: Immediately initiates multi-threaded background compilation of deployed bytecode, parallelizing function-level compilation.
  \item \textbf{Dynamic Hot-Switch Execution}: Provides \textbf{F}unction \textbf{L}evel f\textbf{A}st \textbf{T}ranspile (FLAT) mode and \textbf{F}unction \textbf{L}evel \textbf{A}daptive hot-\textbf{S}witching (FLAS) mode backend compilation services. Upon contract invocation, the engine selects the fastest executable path (FLAT for unoptimized code, FLAS for optimized code) based on compilation progress, ensuring execution never stalls.
\end{itemize}
\textbf{VM Runtime}: The VM Runtime offers pluggable runtime designs for different virtual machine architectures, enabling optimal configurations for JIT-compiled execution. It directly manages the execution lifecycle, including state management, resource control, and enforcement of DTVM's deterministic semantics. 
Take wasm runtime as an example, the key responsibilities include:
\begin{itemize}[itemindent=0.1em]
\item \textbf{Execution Context Management}: Maintains runtime states for each contract instance, including pointers to Wasm linear memory, global variables, and function tables. Enforces dWasm-compliant stack depth and size limits (Section \ref{sec:constrains_and_implementation_for_deterministic_execution}).
\item \textbf{Memory Interface}: Standardizes interactions between JIT-compiled code and Wasm linear memory (with hardware-accelerated bounds checking, Section \ref{sec:efficient_boundary_checking_mechanism}).
\item \textbf{Trap Handling}: Systematically captures and resolves runtime traps (e.g., memory overflows, division-by-zero, stack overflow, invalid indirect calls, gas exhaustion, Section \ref{sec:gas_metering_in_JIT_compilation_mode}–\ref{sec:optimized_JIT_design}).
\item \textbf{Host API Invocation (via VNI Interface)}:
1) Binds imported Wasm functions to native Host implementations registered via VNI (VM Native Interface) modules (e.g., via \lstinline|FUNCTION_LIST|, \lstinline|NATIVE_FUNC_ENTRY| macros).
2) Manages parameter encoding/decoding, context switching, and VM state access via \lstinline|Instance*| pointers.
3) Synchronizes gas consumption and processes return values or exceptions from VNI functions.
4) Manages environment contexts for VNI dependencies (e.g., \lstinline|vnmi_init_ctx|, \lstinline|vnmi_destroy_ctx|).
\item \textbf{Instance Lifecycle Management}: Oversees creation, initialization (e.g., executing start functions), execution, and resource cleanup for Wasm instances.
\end{itemize}

Beyond the above example, other VM runtime such as EVM runtime, can also benifit from this mechanism by adopting EVM specific runtime implementation.

The VM Runtime ensures deterministic execution by tightly integrating with JIT-compiled code, enforcing strict compliance with DTVM’s operational constraints and security requirements. 
\subsection{DTVM SDK}
\label{sec:dtvm_sdk}
DTVM’s adaptation layer supports multi-language contract development, providing frontend design capabilities for contracts written in Solidity, C++, Rust, Go, Java, AssemblyScript, and more. 
It includes SDKs with diverse template libraries to streamline development. 
The core objective is to offer developers a high-efficiency, user-friendly, and multi-language environment that seamlessly integrates with the underlying DTVM engine. 
By providing language-specific SDKs, developers can select the most suitable language based on project requirements, team expertise, or language advantages (e.g., Rust’s security, C++’s performance, or Solidity’s Ethereum compatibility), 
thereby lowering barriers to entry for the DTVM ecosystem and enabling the creation of complex, high-performance decentralized applications. Key features include:

For each supported programming language, we provide a dedicated Software Development Kit (SDK). These SDKs are designed not just as simple interface wrappers but come equipped with comprehensive functionalities to enhance developers' experience when working with the DTVM engine. Core Features of the SDKs Include:

\textbf{Runtime Libraries}: Comprehensive APIs that allow interaction with fundamental DTVM engine capabilities. This includes:
\textit{State Management}, Functions for reading from and writing to contract storage, such as simulating Solidity’s mapping and state variables.
\textit{Context Access}, Easy retrieval of transaction data like sender address (msg.sender) and value (msg.value), along with block information including timestamps and block numbers.
\textit{Event Triggering}, Standardized interfaces for emitting events, facilitating off-chain applications to monitor these events effectively.
\textit{Host API Invocation}, Encapsulated interfaces for calling deterministic Host functions of the DTVM, streamlining complex operations.
\textit{ABI Encoding/Decoding},Tools to process function parameters and return values according to standard interface specifications, ensuring seamless communication between contracts.

\textbf{Build Tool Integration}: The SDKs include plugins or scripts tailored for integration with language-specific build tools such as Maven or Foundry. This integration simplifies the workflow from compiling, optimizing, to packaging source code into executable Wasm bytecode for the DTVM, enhancing development efficiency.

We also provide thorough documentation aimed at guiding developers through the installation and usage of our SDKs, alongside deep dives into core API concepts. To further accelerate onboarding and facilitate understanding, extensive example codes are available, demonstrating practical implementations and best practices.

This approach ensures that developers have all the necessary resources at their disposal to efficiently develop, test, and deploy applications on the DTVM platform.

\subsection{DTVM SmartCogent}
\label{sec:dtvm_tools}
DTVM SmartCogent offers an AI-driven full-stack development experience, leveraging large language models (LLMs) to automate tasks across the smart contract lifecycle: development, debugging, security auditing, and deployment. 
This is achieved through a multi-Agent workflow with serveral defined roles, supported by at least seven specialized Agents.

\begin{figure}[h!] 
  \centering
  \includegraphics[scale=0.5]{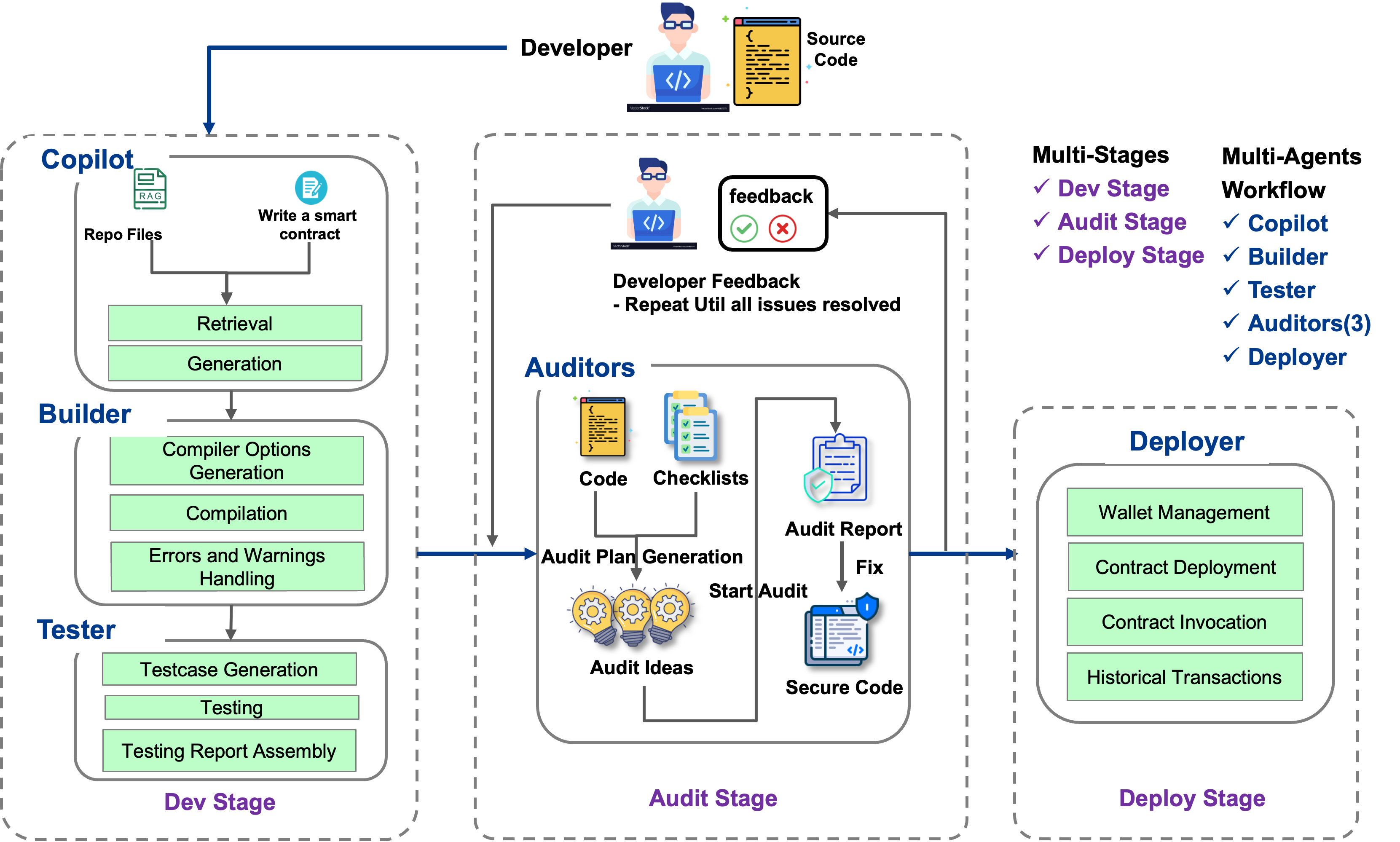}
  \caption{SmartCogent for smart contract development, auditting and deployment}
\end{figure}

\textbf{Contract Development Phase}: Contract development phase consists of three agents,
\textit{Copilot Agent}: Generates code suggestions during development.
\textit{Builder Agent}: Integrates DTVM SDKs via the MCP protocol~\cite{MCP}, automatically detects the programming language, and invokes corresponding toolchains for compilation.
\textit{Tester Agent}: Automates test case generation, execution, and error resolution during testing.

\textbf{Contract Audit Phase}: Building on offline-audited strategies, vulnerability databases, and RAG (Retrieval-Augmented Generation) techniques, DTVM enhances audit effectiveness. The workflow includes:
\textit{Audit Plan Agent}: Generates audit plans tailored to contract characteristics.
\textit{Audit Idea Agent}: Proposes audit strategies based on contract features.
\textit{Audit Report Agent}: Executes audit strategies, generates reports, and suggests code fixes.
\textit{Adversarial Agents}: Simulate malicious behaviors to identify vulnerabilities. 
Through fine-tuned models and RAG-enabled Agents, audit accuracy surpasses that of general-purpose models.

\textbf{Contract Deployment Phase}: Contract deployment phase consists of an \textit{Deployer Agent}, which automates contract deployment and transaction construction via the MCP protocol, interfacing with wallets or blockchain tools to streamline on-chain execution.
This modular Agent architecture ensures seamless, AI-augmented collaboration across all stages of smart contract development and operation.
\section{Technical Implementation}
\label{sec:technical_impl}
\subsection{Implementation of the JIT Engine}
\label{sec:impl_of_JIT_engine}

\begin{figure}[h!] 
  \centering
  \includegraphics[scale=0.5]{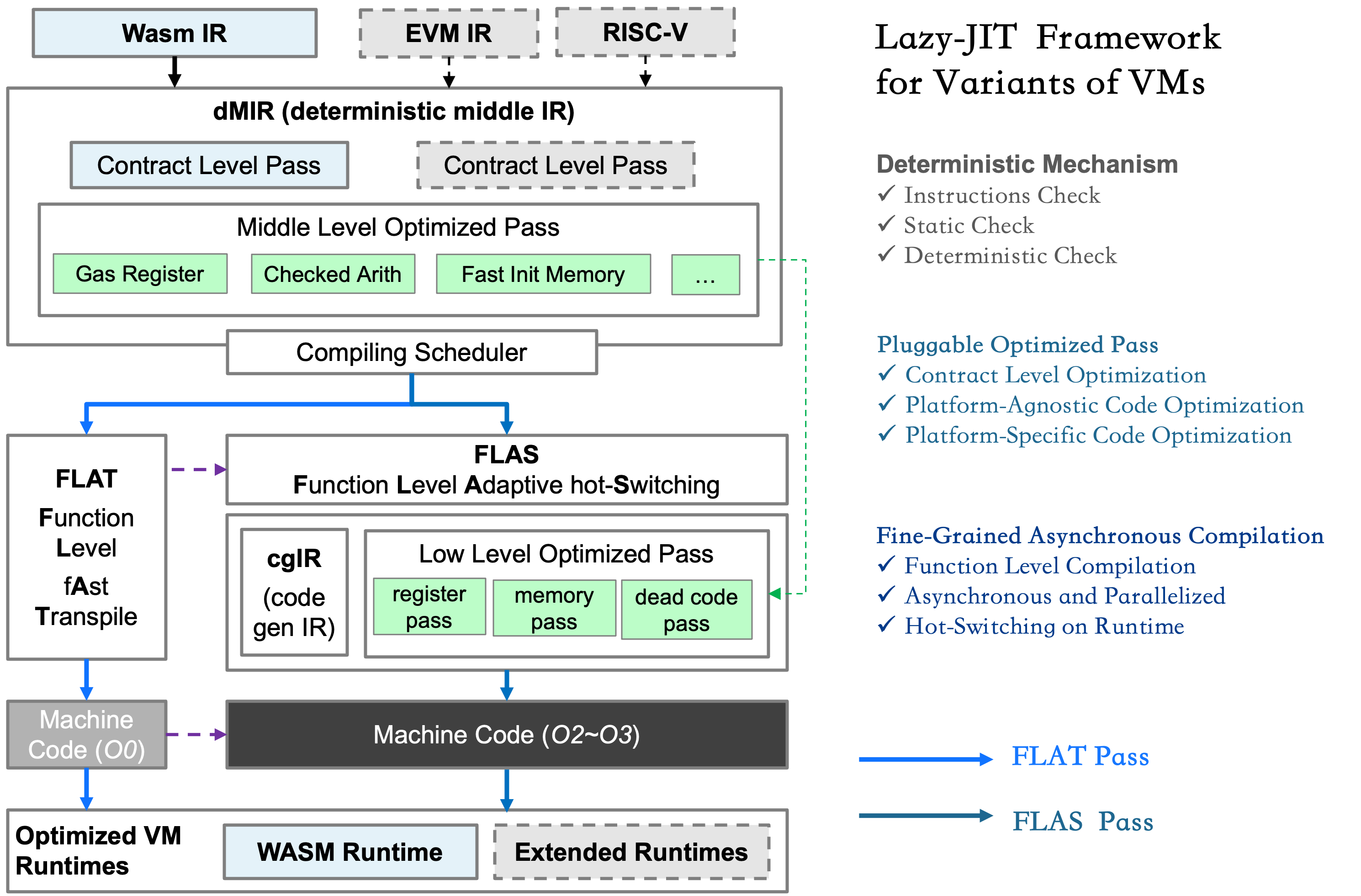}
  \caption{DTVM Lazy-JIT Framework}
\end{figure}

DTVM employs a multi-pass JIT compilation architecture with lazy compilation support to balance the immediate usability of deployed smart contracts with high-performance execution requirements. 
Its core compilation pipeline, illustrated in Figure 3, consists of multiple intermediate representation (IR) layers that progressively lower and optimize input WebAssembly (Wasm) bytecode into target machine code. Compilation workflow and IR layers includes:

\textbf{Input and Initial Processing}: The compilation process begins with the receipt of Wasm bytecode. The first step involves decoding and format validation to ensure compliance with DTVM’s specifications.

\textbf{dMIR Layer (Deterministic Middle IR)}: After validation, WebAssembly (Wasm) instructions are converted into a Deterministic Middle Intermediate Representation (dMIR), which serves as the first intermediate representation (IR) layer in the compilation pipeline. The dMIR layer plays a crucial role in enforcing determinism by ensuring that only modules compliant with the deterministic Wasm (dWasm) specification proceed to subsequent compilation stages, thus guaranteeing deterministic execution from the outset. As a platform-independent IR, dMIR simplifies complex Wasm instructions (such as loops, table operations, and host API calls) into more fundamental operations (computation, memory access, and control flow jumps). This simplification lays a solid foundation for unified processing and optimization across various virtual machine IRs (such as EVM bytecode). During dMIR generation, the system rigorously performs determinism checks, allowing only those Wasm modules with guaranteed deterministic execution to successfully transition into dMIR form. Operations potentially introducing nondeterminism are verified or transformed early in this process. Additionally, the dMIR layer supports a suite of platform-independent optimization passes, such as contract logic optimizations. These optimizations are designed as pluggable modules, offering excellent extensibility and further improving efficiency without compromising deterministic execution.

\textbf{CGIR Layer (Codegen IR)}: Following optimization at the dMIR level, the intermediate representation is further lowered into the Code Generation Intermediate Representation (CGIR). Unlike the platform-independent dMIR, CGIR is platform-specific and tailored to integrate computational tasks closely with the characteristics of the target hardware architecture. At the CGIR layer, the system first performs instruction selection, mapping platform-agnostic dMIR instructions to specific instructions or instruction sequences of the target CPU architecture. Subsequently, register allocation is conducted, mapping an unlimited number of virtual registers onto a limited set of physical registers. DTVM provides multiple strategies for register allocation, including a Greedy Register Allocation (Greedy RA) strategy aimed at peak execution performance and a Fast Register Allocation (Fast RA) strategy designed to balance compilation speed and runtime performance, particularly suitable for rapid compilation during first-time function invocation. Furthermore, the CGIR layer enables CPU architecture-specific optimizations, such as instruction scheduling and peephole optimization. Notably, CGIR may simultaneously contain virtual registers, physical registers, and pseudo-assembly instructions throughout the optimization process, providing comprehensive information required for final code generation.

\textbf{Machine IR Layer and Code Generation}: The optimized CGIR is finally lowered to the Machine IR, a layer closely aligned with target machine instructions. It retains detailed register mappings and pseudo-assembly constructs to facilitate architecture-specific instruction encoding, ultimately producing executable binary machine code.

\begin{lstlisting}[caption = {Example of dMIR (Simplified)}, label=code:example_of_dMIR]
// Function signature: i64 func(i64, i64)
func %func_id (param $arg0 i64, param $arg1 i64) -> i64 {
    var $temp_result i64  // Declare local variable
@block_0:              // Basic block label
    $temp_result = add i64 ($arg0, $arg1) // Basic computation
    return $temp_result                 // Return value
}
\end{lstlisting}

\begin{lstlisting}[caption = {Example of CGIR (Post-Register Allocation for x86\_64)}, label=code:example_of_CGIR]
cgfunc %func_id (reg RCX i64, reg RAX i64) -> reg RAX i64 {
@block_0:
ADD64rr RAX, RCX, RAX  // Platform-specific instruction using physical registers
RET                    // Return instruction
}
\end{lstlisting}

This layered architecture ensures a systematic progression from high-level abstractions to platform-optimized machine code while maintaining deterministic guarantees and performance efficiency.

\textbf{Extensible ISA Support Leveraging Unifide JIT Engine}

DTVM provides a high-performance deterministic JIT engine that compiles contract bytecode IR into dMIR before proceeding with JIT compilation and execution. While current implementations achieve end-to-end functionality through Wasm support, other instruction architectures such as EVM and RISC-V can leverage this system by adding instruction adaptation layers to convert their bytecode into dMIR. This allows them to inherit the performance advantages of DTVM's deterministic JIT engine.

Specifically, DTVM employs a multi-layered architecture design to achieve high modularity and extensibility.
\textit{Frontend Parsing Layer}: Processes smart contract bytecode in diverse formats (Wasm, EVM, RISC-V) and converts it into unified dMIR representation.
\textit{dMIR Optimization Layer}: Performs deterministic optimization on intermediate representations while ensuring execution consistency across heterogeneous environments.
\textit{JIT Compilation Backend}: Efficiently compiles optimized dMIR into native machine code while maintaining deterministic guarantees.
\textit{Runtime Management}: Provides unified memory models, contract interoperability, and resource management mechanisms.

For instance, when handling EVM IR, the primary challenge lies in efficiently mapping u256 data types to physical CPU instruction architectures. This challenge can be addressed by extending support for wide-bit data types at the dMIR layer. For RISC-V instruction architectures, the conversion process demonstrates higher efficiency. As RISC-V natively aligns with mainstream CPU architectures in data bit-width, DTVM's JIT mechanism can generate compact and efficient machine code with optimized translation paths.

\subsection{Lazy Compilation and Stub Design}
\label{sec:lazy_compilation_and_stub_design}
DTVM employs an efficient lazy compilation strategy paired with a meticulously designed stub mechanism to achieve function-level asynchronous JIT compilation. 
This design mitigates the inherent trade-off between compilation time and execution efficiency in smart contract JIT execution, thereby enhancing user experience and system throughput.  

\textbf{Asynchronous Parallel Compilation}

Lazy compilation employs a "compile-on-demand" principle and asynchronous compilation methodology, divided into two core stages: Initially, DTVM quickly parses, validates, and generates dMIR representation upon receiving a Wasm module, creating lightweight stub functions for immediate invocation, while subsequently, a background thread pool conducts multi-pass optimization compilation for individual Wasm functions where compilation priority is determined by strategies such as call frequency prediction or function complexity, generating optimized machine code in parallel for each function to balance startup performance with long-term execution efficiency.

\textbf{Function-Level Stub Design}

Stub functions serve as critical placeholders for uncompiled Wasm functions, guiding execution flow into compilation or lookup processes when invoked before optimization is complete.  

\textit{Structural Design}: 
Each stub typically contains minimal machine instructions. On \textit{x86\_64} architectures, a canonical stub includes a single call instruction redirecting execution to a predefined trampoline (or resolver) function.  
\begin{lstlisting}[caption = {An Example of stub function(x86\_64)}, label=code:example_of_stub_function]
.globl stubTemplatePatchPoint  
stubTemplatePatchPoint:  
    callq stubResolver  // Jump to the resolver function  

.globl stubTemplateEnd  
stubTemplateEnd:  
\end{lstlisting}

\textit{Replaceability}:
Stub functions are dynamically updatable. The target address of the callq instruction serves as a "patch point." Once optimized code is ready, the stub’s jump address is atomically replaced with the new code entry point.  

\textbf{Trampoline Hot-Switch Mechanism}

The Trampoline mechanism serves as the core coordinator for handling scenarios involving the first-time invocation of uncompiled functions, orchestrating a rigorous and efficient workflow. When a Wasm function that has not yet been optimized is invoked through its stub, execution flow jumps to the \textit{stubResolver}, initiating the entire compilation and redirection process. The \textit{stubResolver} first ensures safety by saving the state of all general-purpose registers and using the textit{fxsave64} instruction to preserve the floating-point/vector register state onto the stack, thus protecting the original execution context from disruption during subsequent compilation steps. Next, the \textit{stubResolver} retrieves the return address from the stack as the second parameter and retains the Wasm instance pointer as the first parameter. It then invokes the \textit{compileOnRequestTrampoline} function via an instruction at the \textit{stubResolverPatchPoint}. This function identifies the Wasm function requiring compilation, performs the necessary compilation tasks, and returns the entry address of the newly compiled code (optimized). Upon receiving the compiled function's address, the \textit{stubResolver} overwrites the original return address location on the stack, effectively redirecting the execution flow. After updating the address, the \textit{stubResolver} restores the floating-point/vector register state using the \textit{fxrstor64} instruction and recovers all previously saved general-purpose registers. Finally, the \textit{stubResolver} executes a \textit{ret} instruction; however, as the return address on the stack has been modified, execution flow directly jumps into the newly compiled function's entry point instead of returning to the original stub code. This seamlessly transitions execution from an uncompiled state to optimized code execution.

\textit{Note}: The stub update is performed internally within compileOnRequestTrampoline, using the provided stub address to atomically modify the patch point.  
This architecture ensures seamless execution flow transitions between unoptimized stubs and optimized code while maintaining deterministic guarantees and minimal runtime overhead.

\textbf{Thread Safety Considerations}

DTVM's lazy compilation design carefully addresses concurrency issues arising in multi-threaded environments, ensuring system stability and efficiency under high concurrency scenarios. When multiple threads simultaneously invoke the same uncompiled function stub, they may concurrently enter the Trampoline mechanism, potentially performing compilation tasks in parallel. However, upon completion of compilation, the system employs atomic operations to update the stub, ensuring all threads consistently observe the latest compilation results. Notably, even in scenarios where multiple threads concurrently compile the same function, the idempotent nature of the compilation results guarantees correctness, with only minimal additional resource usage. Additionally, DTVM meticulously handles synchronization between background compilation tasks and the lifecycle of Wasm runtime instances. The system ensures safe cancellation or waiting for associated background compilation tasks upon runtime release, effectively preventing dangling pointers or resource contention. Moreover, it guarantees that runtime instances remain valid when accessed by background compilation threads, thus maintaining overall system stability and data consistency.

\begin{minipage}{\linewidth}
\begin{lstlisting}[caption = {StubResolver Assembly Example (x86\_64)}, label=code:stubresolver_assembly_example, breaklines=false]
.globl stubResolver
.type  stubResolver, @function
stubResolver:
/* can't add breakpoints in assembly templates,
* otherwise the breakpoint postions will be changed
* to 0xcc(int3) by debugger */
/* save registers */
push %rbp
mov %rsp, %rbp
push %rax
push %rbx
push %rcx
push %rdx
push %rdi
push %rsi
push %r8
push %r9
push %r10
push %r11
push %r12
push %r13
push %r14
push %r15
push %rsp
/* and %rsp, -128 */ /* make rsp 128bytes aligned */
/* allocate 560(at least 256 and rsp need 128 aligned) bytes to save all registers */
sub $560, %rsp
/*fxsave64 arg address must 128-aligned, XMM-s saved in +128 positions, MXCSR saved in +256 postions*/
/* save registers */
fxsave64 (%rsp)
/* get trampolineAddr as compileOnRequestTrampoline second argument */
/* when trampoline call resolver, the trampoline rip will saved in stack */
movq 0x08(%rbp), %rsi
/* DTVM will replace the trampoline address here, to allow stubResolver call it */
/* the label stubResolverPatchPoint is used to find the memory position to update */
.globl stubResolverPatchPoint
stubResolverPatchPoint:
/* moveabsq instructions has 10bytes(last 8bytes is i64 value) */
movabsq $0, %rax /* this target address need updated to compileOnRequestTrampoline */
/* first argument register not changed, the first still WASMInstance* */
/*compileOnRequestTrampoline pointer should call by abosulte because trampoline codo generated dynamic*/
call *%rax
/* change stack to override the next instruction after ret. this works like re-entry */
mov %rax, 8(%rbp)
/* restore registers from memory */
fxrstor64 (%rsp)
add $560, %rsp
pop %rsp
pop %r15
pop %r14
pop %r13
pop %r12
pop %r11
pop %r10
pop %r9
pop %r8
pop %rsi
pop %rdi
pop %rdx
pop %rcx
pop %rbx
pop %rax
mov %rbp, %rsp
pop %rbp
ret
.globl stubResolverEnd
stubResolverEnd:
\end{lstlisting}
\end{minipage}

\subsection{Efficient Boundary Checking Mechanism}
\label{sec:efficient_boundary_checking_mechanism}
A critical requirement for a VM as an execution sandbox is ensuring memory access safety to prevent out-of-bounds vulnerabilities or system crashes. 
Traditional boundary checking mechanisms rely on explicit software checks, which are reliable but incur significant performance overhead. 
DTVM leverages modern processor memory protection mechanisms to implement a nearly \textbf{zero-overhead} boundary checking scheme, significantly improving execution efficiency while maintaining isolation.  

\textbf{Traditional Boundary Checking Approach}: 
In conventional WebAssembly VM implementations, memory access is guarded by explicit checks, such as:  
\begin{lstlisting}[label=code:traditional_boundary_checking_approach]
// Pseudo-code for traditional explicit checking  
if (address + offset >= memory_size) {  
  throw OutOfBoundsException();  
}  
value = memory[address + offset];  
\end{lstlisting}

Performance Bottlenecks of Traditional Methods:   
\textit{Instruction Overhead}: Each memory access requires additional comparison and conditional jump instructions.  
\textit{Branch Prediction Pressure}: Frequent conditional checks strain CPU branch predictors.  
\textit{Instruction Cache Contamination}: Redundant check code occupies limited instruction cache space.  
\textit{Pipeline Stalls}: Conditional jumps may trigger CPU pipeline flushes.

These overheads accumulate into significant performance bottlenecks in memory-intensive smart contract execution scenarios, particularly in compute-heavy workflows.  

DTVM’s boundary checking mechanism avoids these issues by offloading checks to hardware-level memory protection (e.g., page-based access control), 
eliminating redundant software checks while maintaining safety guarantees. 
This approach aligns with modern CPU architectures, balancing performance and security without sacrificing determinism.  

\textbf{Hardware-Based Boundary Checking Mechanism}

DTVM implements a zero-overhead boundary checking mechanism leveraging hardware-based memory protection and CPU exception handling. 
This design offloads traditional software checks to the underlying hardware, ensuring both safety and performance.  

\begin{figure}[h!] 
  \centering
  \includegraphics[scale=0.5]{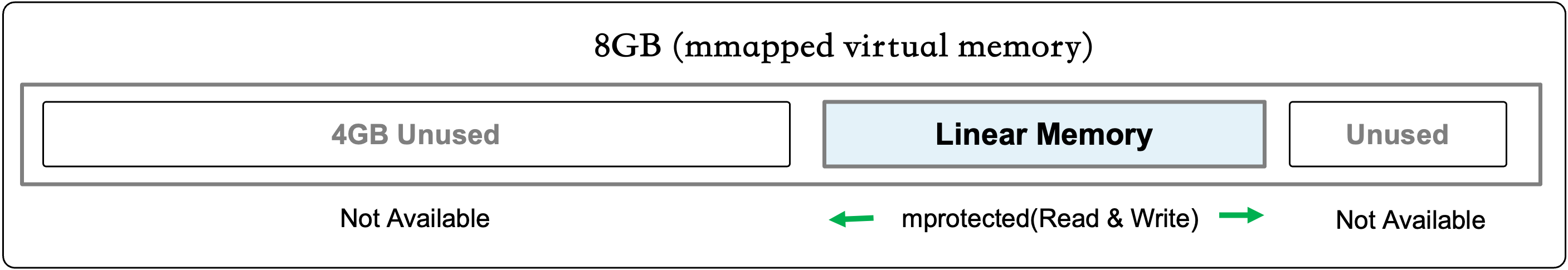}
  \caption{Layout of virtual memeory}
  \vspace{16pt}
\end{figure}

\textbf{Virtual Memory Layout Strategy}: The core of DTVM’s memory protection relies on a carefully designed virtual memory mapping structure: 

\begin{lstlisting}[label=code:example_of_CGIR]
void* base = mmap(  
0,  
8 * 1024 * 1024 * 1024,  // 8GB  
PROT_NONE,               // Initially inaccessible  
MAP_PRIVATE | MAP_ANONYMOUS,  
-1, 0  
);  
\end{lstlisting}

\textbf{Large Address Space Reservation}: An 8GB virtual address space is reserved using the mmap system call, spanning from the linear memory base address -4GB to +4GB. This range aligns with the theoretical addressable bounds of 32-bit WebAssembly memory access instructions. 

\textbf{Active Region Protection}: Only the actual memory region in use is granted read/write permissions via \lstinline|mprotect|:  

\begin{lstlisting}[label=code:active_region_protection]
mprotect(  
actual_memory_base,      // Start of linear memory  
actual_memory_size,      // Current allocated size  
PROT_READ | PROT_WRITE   // Enable access  
);  
\end{lstlisting}

\textbf{Guarded Isolation}: The regions outside the active memory area within the reserved 8GB remain marked as \lstinline|PROT_NONE|, acting as "guard pages." Any access to these pages triggers a hardware exception.
This design ensures that:  
1) WebAssembly’s 32-bit signed base address plus 32-bit unsigned offset calculations remain within the reserved address space.  
2) Out-of-bounds accesses (e.g., exceeding the linear memory size) map to guard pages, 
which generate hardware exceptions instead of accessing unintended memory regions.

\textbf{JIT Compilation Optimizations}: Building on this memory layout, DTVM's JIT compiler removes explicit software checks and generates direct memory access instructions:  

\begin{lstlisting}[label=code:JIT_compilation_optimizations]
; Optimized memory load (x86-64 pseudo-assembly)  
; rdi = base address, rsi = offset, rax = result register  
lea rax, [rdi + rsi]      ; Compute effective address  
mov eax, [rax]            ; Direct access-no explicit bounds check
; Hardware traps if [rax] points to a guard page  
\end{lstlisting}

\textbf{Exception Handling Mechanism}: 
When an out-of-bounds memory access occurs (such as accessing a protected memory page), CPU hardware generates a segmentation fault or page fault, typically manifested as a \textit{SIGSEGV} signal on Unix-like systems. To handle these hardware exceptions, DTVM implements a comprehensive exception handling mechanism by registering custom signal handlers.
DTVM first registers a handler for the \textit{SIGSEGV} signal using the \textit{sigaction} system call, setting the \textit{SA\_SIGINFO} flag to obtain detailed fault information, including the specific memory address causing the fault. The registration code is as follows:

\begin{lstlisting}[label=code:exception_handling_mechanism]
struct sigaction sa;
memset(&sa, 0, sizeof(sa));
sa.sa_sigaction = dtvm_sigsegv_handler; // Custom handler function
sa.sa_flags = SA_SIGINFO | SA_NODEFER; // Get fault info, allow recursive signals if needed
sigaction(SIGSEGV, &sa, NULL);
\end{lstlisting}

When a \textit{SIGSEGV} signal is triggered, the \textit{dtvm\_sigsegv\_handler} carefully analyzes the signal's context information, checking whether the instruction pointer (\textit{rip} or \textit{pc}) resides within the JIT-generated Wasm code region and confirming the faulting memory address (\textit{siginfo\_t->si\_addr}). The handler then determines whether the faulting address falls outside the valid range of the current Wasm instance's linear memory but within its reserved virtual address space.
If confirmed as a Wasm linear memory out-of-bounds access, DTVM gracefully converts this hardware exception into a WebAssembly-defined trap, such as an exception with an "out of bounds memory access" message. This conversion typically involves setting flags or using mechanisms like \textit{longjmp} or exception throwing to safely transfer control back to the VM's exception handling logic.
For \textit{SIGSEGV} signals caused by non-Wasm code, the system restores default signal handling behavior or forwards the signal to previously registered handlers, ensuring proper handling of other segmentation fault types and maintaining overall system stability and reliability.

\textbf{Memory Growth and Multi-Instance Isolation}: 
\textit{Memory Expansion}: When expanding linear memory (e.g., via memory.grow), DTVM simply adjusts the \lstinline|PROT_READ||\lstinline|PROT_WRITE| region within the reserved space using mprotect, avoiding costly reallocations.  
\textit{Instance Isolation}: Each Wasm instance operates within its own 8GB virtual address space. This ensures that out-of-bounds accesses in one instance do not interfere with others, even in parallel execution scenarios (e.g., concurrent smart contract execution).

\textbf{Performance and Safety Trade-off}: 
By leveraging hardware-based memory protection, DTVM eliminates software-boundary-check overhead entirely. This approach ensures:  
1) Zero runtime cost for valid memory accesses, 
2) Deterministic safety via hardware-enforced isolation, 
3) Scalability for compute-intensive or memory-heavy smart contracts.
This mechanism aligns with modern CPU architectures, delivering both security guarantees and performance efficiency critical for high-throughput decentralized applications.

\subsection{Gas Metering in JIT Compilation Mode}
\label{sec:gas_metering_in_JIT_compilation_mode}
In blockchain scenarios, the gas mechanism is a foundational infrastructure for managing computational resource consumption, 
preventing infinite loop attacks, and ensuring network fairness. 
Each operation in smart contracts consumes a predefined gas quantity, requiring execution to terminate before pre-paid gas is exhausted. 
Traditional interpreter-based virtual machines typically perform gas checks and deductions within interpreter loops or via instrumentation of gas-counting instructions, which incur significant instruction and memory overhead. 
DTVM innovatively integrates gas metering with JIT compilation, designing efficient strategies tailored to different optimization levels while maintaining low overhead, precision, and determinism. Key challenges include:  
\textit{Performance Overhead}: Frequent gas deductions and checks consume computational resources, degrading execution efficiency.  
\textit{Precision and Determinism}: gas consumption must be identical across all nodes and accurately reflect computational costs.  
\textit{JIT Integration}: gas logic must be embedded into optimized machine code generation without compromising correctness or performance.

DTVM adopts distinct gas metering implementations for its JIT compilation modes (FLAT vs. FLAS) to balance compilation speed, optimization granularity, and runtime costs. 
The core approach involves inserting gas deduction and check logic at basic block entry points or around specific instructions (e.g., loops, function calls).  

\textbf{FLAT Mode (Dedicated Physical Register Scheme)}: 
In the FLAT mode, which emphasizes rapid compilation, DTVM employs an efficient gas metering strategy using a dedicated physical register. During compilation, a general-purpose physical register (e.g., \textit{rbx} on \textit{x86\_64} or \textit{x22} on \textit{ARM64}, typically not used for parameter passing or return values) is reserved exclusively for tracking the remaining gas.
During JIT compilation, the system calculates the gas cost for each basic block and directly generates corresponding machine instructions at the block's entry point to decrement the gas register (e.g., \textit{sub\ rbx, cost}) and check for gas exhaustion (e.g., \textit{jb out\_of\_gas\_handler}). Additionally, other computation instructions generated by the JIT avoid using this dedicated gas register to prevent conflicts.

\textbf{FLAS Mode (Virtual Register-Based Scheme)}: 
In the FLAS mode that focuses on deep optimization, DTVM adopts a virtual-register based gas metering scheme, which provides greater flexibility and optimization space. In this scheme, the gas counter is represented as a virtual register at the dMIR level, while gas deduction and checking logic are inserted as IR instructions at the entrance of basic blocks in dMIR instructions.
Although the number of instructions for a single check may be more than FLAT (if the gas value is allocated to memory), overall better performance can be achieved through compiler optimization, especially in complex functions, where register allocation is more flexible and can more fully utilize register resources to reduce overall memory access.

\textbf{Host API Interaction Handling}: 
Regardless of compilation mode, gas metering must synchronize with Host API calls to ensure continuity and accuracy.
\textit{Pre-Call Save}: Before invoking Host APIs, the current gas value (stored in a dedicated register or spilled memory) is saved to the Wasm instance’s memory state.  
\textit{Host Execution}: Host APIs may consume gas (via internal logic or external frameworks), updating the instance’s gas value in memory.  
\textit{Post-Return Sync}: After returning from Host APIs, the gas value is reloaded into the VM’s gas counter (register or virtual register) to maintain synchronization.  
\textit{Exception Handling}: gas state is preserved and restored if some Wasm traps happen.

DTVM’s gas metering strategy achieves low overhead and deterministic behavior through several ways.
\textit{Mode-Specific Optimization}: 
1) FLAT Mode: Minimal per-check overhead for rapid execution, 
2) FLAS Mode: Compiler-driven optimizations reduce aggregate memory access costs in complex workloads.
\textit{Hardware-Level Synchronization}: Dedicated registers or spill mechanisms ensure gas values remain consistent across all execution paths.  
\textit{Host API Isolation}: Explicit save/restore protocols guarantee gas state integrity during external interactions.

By tightly integrating gas metering with JIT compilation and leveraging mode-specific optimizations, DTVM ensures precise resource control without compromising performance, critical for high-throughput blockchain applications.  This design aligns with hardware constraints and compiler optimizations, delivering deterministic gas accounting while minimizing runtime overhead.

\subsection{Optimized JIT Design for Integer Overflow Checks in Smart Contracts}
\label{sec:optimized_JIT_design}
Integer overflow vulnerabilities are a common and critically dangerous issue in smart contracts, historically leading to substantial financial losses. 
Standard mitigation approaches, such as inserting explicit overflow checks after arithmetic operations (e.g., addition, multiplication) or relying on compiler-enforced safe operations (e.g., Solidity 0.8+), often incur significant performance penalties and increase computational costs. 
DTVM addresses this challenge through a novel JIT optimization technique that efficiently handles overflow checks while minimizing performance degradation. 

Traditional overflow check mechanism has limitations.
\textit{Performance Overhead}: Requires additional comparison and conditional jump instructions.  
\textit{Code Redundancy}: Manual checks must be duplicated across all arithmetic operations.  
\textit{Prone to Omissions}: Vulnerable to human error in omitting checks for certain operations.

An explicit code check example is listed below.

\begin{lstlisting}[caption = {Example (Solidity < 0.8 or manual checks)}, label=code:example_of_solidity]
function safeAdd(uint256 a, uint256 b) internal pure returns (uint256) {  
  uint256 c = a + b;  
  require(c >= a, "SafeMath: addition overflow"); // Explicit check  
  return c;  
}  
\end{lstlisting}

Leveraging compiler-provided functions or instructions to complete the check is straight forword. However, it also suffers from some drawbacks.  
\textit{Instruction Bloat}: Even simple overflow checks generate multiple instructions (e.g., function calls, result extraction, conditional branches).  
\textit{Code Size and Execution Time}: Increases binary size and execution latency, potentially hindering further optimizations.  
\textit{Wasm Translation Penalty}: When compiled to WebAssembly, these checks result in verbose Wasm instructions, exacerbating performance overhead. The example case is shown as below.

\vspace{16pt}

\begin{minipage}{\linewidth}
\begin{lstlisting}[caption = { Example (C → LLVM IR):}, label=code:example_of_c_to_llvm_ir]
int a, b; // params
int res;
if (__builtin_add_overflow(a, b, &res)) {
    panic();
}

// generated llvm IR
%result_pair = call {i32, i1} @llvm.sadd.with.overflow.i32(i32 %a, i32 %b)
%result = extractvalue {i32, i1} %result_pair, 0
%overflow_bit = extractvalue {i32, i1} %result_pair, 1
br i1 %overflow_bit, label %overflow_handler, label %continue
\end{lstlisting}
\end{minipage}

An optimized strategy is proposed in this work, which is reffered as "\textit{Hook Functions and JIT Inline Optimization}". DTVM employs a "semantic tagging and JIT specialization" approach to address integer overflow checks efficiently:  

\textbf{Defining Overflow Check Hook Functions}: 
A set of specially named imported functions (hook functions) is defined at the Wasm IR level to semantically mark arithmetic operations requiring overflow checks. These functions act as markers for the JIT compiler rather than containing complex implementations.
\textit{Naming Convention}: Follows \lstinline|checked_{type}_{operation}| (e.g., \lstinline|checked_i32_add|, \lstinline|checked_u64_mul|), where \lstinline|{type}| specifies the integer type (e.g., \lstinline|i32|, \lstinline|u64|) and \lstinline|{operation}| denotes the arithmetic operation (e.g., \lstinline|add|, \lstinline|mul|). 
\textit{128-Bit Integer Handling}: For unsupported 128-bit types, parameters are split into 64-bit chunks (e.g., \lstinline|lo1, hi1, lo2, hi2|). Results are returned via convention (e.g., high 64 bits for \lstinline|checked_u128_add|).

\begin{lstlisting}[caption = {Wasm Module Example}, label=code:wasm_module_example]
(module
    ;; Import hook for signed 32-bit addition with overflow check
    (import "env" "checked_i32_add" (func $checked_i32_add (param i32 i32) (result i32)))
    ;; Import hook for unsigned 64-bit multiplication with overflow check
    (import "env" "checked_u64_mul" (func $checked_u64_mul (param i64 i64) (result i64)))
    ;; Import hook for unsigned 128-bit addition (special convention)

    (import "env" "checked_u128_add" (func $checked_u128_add (param i64 i64 i64 i64) (result i64))) 
    ;; lo1, hi1, lo2, hi2 -> hi_result

    (func $example_usage (param $a i32) (param $b i32) (result i32)
    local.get $a
    local.get $b
    call $checked_i32_add ;; Use the hook function
    )
)
\end{lstlisting}

\textbf{JIT Compilation Recognition and Specialization}:
During compilation, the JIT compiler identifies calls to hook functions and performs inline optimization. When a hook function is detected (e.g., \lstinline|checked_i32_add|), the JIT compiler directly generates optimized CPU-specific instructions at the call site instead of emitting a function call via inline machine code.  
Meanwhile, CPU arithmetic flags are leveraged to perform overflow checks with minimal overhead.

\textbf{Hardware Overflow Flag Utilization}
Modern CPUs (e.g., x86\_64, ARM64) automatically set status flags during arithmetic operations. 
As for \textit{x86\_64}: 
1) Signed Operations (add, sub, imul): Check the Overflow Flag (OF).   
2) Unsigned Operations (add, sub, mul): Check the Carry Flag (CF). 
As for \textit{ARM64}:  
1) Signed Operations: Use \lstinline|adds/subs| instructions and check the V (overflow) flag.  
2) Unsigned Operations: Check the C (carry) flag.

\textbf{Optimized Machine Code Generation}
The JIT compiler generates concise code sequences for each overflow check.
The sequence includes:  
1) A single arithmetic instruction to compute the result and update flags.  
2) A conditional jump (e.g., \lstinline|jo, jno, jc|) to handle overflow.  
3) Overflow paths trigger a unified trap handler to terminate execution.

\begin{lstlisting}[caption = {Example for an optimized \lstinline|checked_i32_add| (x86\_64)}, label=code:wasm_module_example]
; Input: eax = operand1, ebx = operand2
; Output: eax = result (if no overflow)

add eax, ebx       ; Perform addition, sets OF flag on signed overflow
jo near overflow_trap ; Jump if Overflow Flag (OF) is set
\end{lstlisting}

\textbf{Advantages of the Optimized Approach}
\textit{Performance Improvement}:  
  1) In non-overflow cases, execution requires only one arithmetic instruction and an unused branch prediction, eliminating redundant checks.  
  2) Hardware flag checks incur negligible overhead.
\textit{Code Size Reduction}: Inlined code is more compact than external function calls or expanded check logic.
\textit{Deterministic Behavior}: Overflow triggers a VM trap, halting execution and consuming gas up to the overflow point, ensuring safety and determinism.

This design allows developers to write secure smart contracts (e.g., with overflow-safe arithmetic) without compromising execution speed, aligning with blockchain’s demand for both security and efficiency.

\subsection{Constraints and Implementation for Deterministic Execution}
\label{sec:constrains_and_implementation_for_deterministic_execution}
Deterministic execution is a foundational requirement for blockchain consensus: identical smart contract code and inputs must produce the same outputs, state changes, and gas consumption across all nodes, regardless of CPU architecture, operating system, or virtual machine implementation. 
While standard WebAssembly (Wasm) prioritizes compatibility with browsers and general-purpose computing, it lacks strict guarantees for determinism. 
DTVM addresses this gap by defining and enforcing the \textbf{deterministic WebAssembly (dWasm)} extension specification and runtime constraints, systematically eliminating sources of uncertainty in standard Wasm to ensure deterministic execution for smart contracts.  


Standard Wasm specifications introduce several factors that may lead to non-deterministic behavior across environments or execution instances:  

\textbf{Static/Load-Time Uncertainties}: 
\textit{Validation Rules}: Ambiguities or implementation-defined behaviors in validation rules may lead to inconsistent module acceptance.  
\textit{Error Reporting}: The specific error reported during validation (e.g., format errors, type mismatches) may vary.  
\textit{Resource Limits}: Insufficiently strict or non-standardized constraints on module, function, memory, and table sizes.

\textbf{Execution-Time Uncertainties}: 
\textit{Floating-Point Operations}:  IEEE 754 floating-point arithmetic may produce slight differences across CPU architectures or compiler optimizations (e.g., handling NaNs, subnormal numbers, or rounding modes).  
NaN representations (payload and sign bits) may not be uniquely defined.
\textit{Type Conversions}: Floating-to-integer conversions for out-of-range values or NaNs may depend on platform-specific behavior.
\textit{Host Functions}: Imported functions’ behaviors are entirely defined by the host environment, introducing non-determinism.  
\textit{Memory Operations}: Atomic operation memory model details or undefined behaviors (e.g., unaligned memory accesses) may vary.

\textbf{Exception/Trap Uncertainties}: 
\textit{Stack Overflow}: No standardized stack size limits or overflow detection mechanisms.  
\textit{Resource Exhaustion}: Behavior upon resource depletion (e.g., memory overflow) may differ.  
\textit{Trap Reporting}: Details of trap events (e.g., division by zero, memory out-of-bounds) and termination mechanisms may not be consistent.


DTVM enforces determinism through dWasm constraints across its compiler, intermediate representation (dMIR), and runtime execution layers:  

\textbf{Strict Static Constraints and Validation}: 
During module loading and validation, DTVM applies stricter checks than standard Wasm:  
\textit{Fixed Resource Limits}: 
  1) Function parameter count: $\leq 1024$, 
  2) Function local variable count: $\leq 10,240$ (all types combined), 
  3) Per-function stack frame size (calculated per dWasm rules): $\leq 40,960$ (e.g., $i32/f32 = 1, i64/f64 = 2$), 
  4) Per-function instruction count: $\leq 102,40$, 
  5) Control flow nesting depth (blocks, loops, ifs): $\leq 1,024$, 
  6) Explicit limits on memory pages, table sizes, and import/export counts.
\textit{Deterministic Validation}:  
1) Validation steps are strictly ordered. The first detected dWasm violation triggers an immediate rejection with consistent error reporting across nodes.  
2) Format Normalization: Modules using optional Wasm features or non-zero padding are rejected. UTF-8 identifiers (imports/exports) are validated for strict encoding.  
3) Early Rejection: Static checks and resource limits are enforced early in the compilation pipeline (before dMIR generation). Non-compliant modules are rejected deterministically, independent of JIT compilation mode (SinglePass/MultiPass/Lazy).

\textbf{Deterministic Runtime Environment}: 
\textit{1. Virtual Stack Model}:  
\textit{Fixed Size}: Each Wasm instance uses a fixed-size virtual call stack (e.g., 8MB), independent of the host system.  
\textit{Depth Limits}: Maximum call depth (e.g., 1,024 layers) and stack overflow checks are enforced before each function call. Exceeding limits triggers a deterministic trap.
\textit{2. Host Function Controls}: All external host functions must exhibit deterministic behavior. Non-deterministic operations (e.g., system time, random numbers) are prohibited or replaced with blockchain-specific deterministic inputs (e.g., block header data).
\textit{3. Trap Handling}: 
\textit{Uniform Trap Types}: Defined trap types include unreachable, \lstinline|memory_access_out_of_bounds|, \lstinline|integer_divide_by_zero|, and others.  
\textit{Immediate Termination}: Any trap halts execution immediately.  
\textit{gas Consumption}: Trap gas costs are precisely defined (typically consumed up to the triggering operation).

To ensure determinism across different JIT compilation modes and lazy compilation workflows, 
DTVM relocates all determinism checks from the loading phase to before dMIR generation. 
This ensures that any WebAssembly module incapable of deterministic JIT compilation will fail at the dMIR generation stage rather than during lazy execution. Only modules that successfully generate dMIR proceed to JIT compilation, thereby eliminating uncertainty introduced by deferred compilation.

DTVM enforces determinism through the \textit{dWasm extension specification}, imposing strict constraints across multiple dimensions:
\textit{Static Validation}: Pre-execution checks for module structure and compliance with dWasm constraints.
\textit{Resource Limits}: Enforced memory, stack, and computational bounds to prevent ambiguous states.
\textit{Runtime Behavior}: Strict management of floating-point operations and stack usage to avoid non-deterministic outcomes.
\textit{Error Handling}: Consistent trap mechanisms for invalid operations, ensuring uniform error propagation.

These measures systematically eliminate indeterminacy inherent in standard WebAssembly, guaranteeing that smart contracts executed on DTVM produce \textit{identical results across all compatible nodes}. 
This aligns with blockchain consensus requirements for deterministic execution. 
Determinism checks are deeply integrated into both compilation and runtime workflows, with early-stage validation ensuring compatibility with diverse JIT strategies—including lazy compilation—while preserving consistency.

\subsection{Solidity Compilation Optimization and Multi-Language Support}
DTVM aims to establish an open, developer-friendly ecosystem for smart contracts by supporting multiple mainstream programming languages. This enables developers to leverage familiar tools and languages for WebAssembly-based smart contract development. This section outlines the languages supported by DTVM, their compilation workflows, and the optimization strategies for Solidity—the core language of the Ethereum ecosystem.  
DTVM provides the following language support for compiling to WebAssembly bytecode:  

\textit{C/C++}: 
1) Leverages the mature Clang/LLVM toolchain combined with DTVM’s code generation tools.  
2) Developers define contract interfaces, event declarations, and storage layouts in Solidity, then implement logic in C/C++ by inheriting from Solidity abstract classes.  
3) DTVM provides header files and libraries for blockchain state access, event handling, and host API integration.  
4) Code is compiled to LLVM Intermediate Representation (IR) via Clang and then to Wasm using LLVM’s WebAssembly backend.

\textit{Rust}: 
1) Utilizes Rust’s native WebAssembly support (via the wasm32-unknown-unknown target) through the official Rust compiler (rustc).  
2) Rust’s memory safety guarantees (no garbage collection), strong typing, and low-level control make it ideal for secure, high-performance contracts.  
3) DTVM offers a Rust SDK to abstract interactions with the virtual machine environment.

\textit{Solidity}: Compiles via a Solidity → Yul~\cite{Yul} → WebAssembly pathway, a critical feature for Ethereum ecosystem compatibility (Detailed in section 4.7.1.)

\textit{Java}:  
1) Employs direct conversion from Java bytecode to WebAssembly, bypassing the traditional JVM. 
2) Implements Java features (exceptions, reflection, polymorphism, inheritance, and garbage collection) at compile time, enabling Web2 developers to transition to Wasm contracts with minimal effort.

\textit{Go}: Leverages TinyGo’s WebAssembly support, along with DTVM-provided libraries to assist in developing Go-based smart contracts.

\textit{AssemblyScript}:   
1) A statically typed subset of TypeScript designed for WebAssembly.  
2) Contracts are compiled directly to Wasm using the AssemblyScript compiler (asc), offering a web-development-like experience while generating efficient bytecode.

\textbf{Solidity to Wasm}

To ensure seamless compatibility with the Ethereum ecosystem, DTVM provides robust Solidity support through its core strategy of leveraging the Yul intermediate representation from the Solidity compiler, following a Solidity → Yul IR → wasm compilation workflow with optimizations including type inference and refinement, memory access optimization, function inlining, constant folding/propagation, dead code elimination, and control flow optimization. This approach allows DTVM to reuse solc's mature parsing and type-checking capabilities while focusing on backend optimization with WebAssembly-specific transformations that surpass EVM limitations through WebAssembly's richer instruction set.
DTVM further enhances developer experience through specialized libraries and tooling for each supported language, with features like Solidity interface embedding, storage model mapping, event compatibility APIs, and automated ABI handling that maintains perfect interoperability with Ethereum's contract ABI standards while delivering superior performance through WebAssembly's efficient execution model.

\textbf{Optimizations at the Yul Level:}

\begin{itemize}
    \item[$1.$] \textbf{Type Inference and Refinement}: Yul’s default 256-bit operand width (mimicking EVM behavior) is reduced to smaller native WebAssembly types (e.g., \lstinline|i32|, \lstinline|i64|) through static analysis. For example, operations like selector reads or non-zero checks for call values require only 32 or 64 bits, reducing computational overhead.  
   
    \item[$2.$] \textbf{Memory Access Optimization}:  \textit{1) Memory Mapping}: EVM memory operations (e.g., \lstinline|mstore|) are mapped to WebAssembly’s linear memory, leveraging its fixed offset growth mechanism. \textit{2) Redundancy Elimination}: Removes redundant \lstinline|mload/mstore| operations.
    
    \item[$3.$] \textbf{Function Inlining}: Reduces overhead by inlining small, frequently called Yul helper functions.  

    \item[$4.$] \textbf{Constant Folding/Propagation}: Evaluates constant expressions at compile time and propagates their values to usage points.  

    \item[$5.$] \textbf{Dead Code Elimination}: Removes unreachable code paths.  

    \item[$6.$] \textbf{Control Flow Optimization}: Simplifies jump logic (e.g., \lstinline|switch, if, for|) and merges basic blocks.

\end{itemize}

\textbf{Yul to Wasm Pipeline:}
\begin{itemize}
  \item[$1.$] \textbf{Leverage Official Frontends}: Reuses solc’s mature parsing and type-checking capabilities.  

 \item[$2.$] \textbf{Focus on Backend Optimization}: Enables DTVM to concentrate on Yul-to-Wasm transformations with WebAssembly-specific optimizations.  

 \item[$3.$] \textbf{Surpass EVM Limitations}: WebAssembly’s richer instruction set (e.g., \lstinline|i32/i64| operations) yields more efficient code than equivalent EVM bytecode.
\end{itemize}

\textbf{More programming Language Support}

DTVM extends its multi-language support through comprehensive tooling for traditional programming languages. 
For C/C++, DTVM provides specialized libraries (e.g., contractlib.hpp) that simplify smart contract development through Solidity interface embedding (auto-generating abstract base classes from Solidity interfaces), storage model mapping (with C++ classes like Storage<T> and StorageMap<K,V> that encapsulate Solidity-style storage), event compatibility APIs (generating functions that trigger Solidity ABI-compatible events), and ABI automation (handling encoding/decoding for contract entrypoints). The C/C++ compilation workflow leverages Clang to generate LLVM IR before transformation into Wasm via LLVM's WebAssembly backend. 
For Java developers, DTVM implements advanced static compilation strategies that translate Java's dynamic features into WebAssembly's static environment, including exception handling (converting try-catch constructs into control-flow equivalent code), reflection APIs (transforming analyzable reflection calls into direct method invocations), polymorphism support (optimizing virtual method calls through type hierarchy analysis), deterministic garbage collection (tailored for smart contract execution), and preservation of generic type information despite Java's type erasure. These comprehensive adaptations ensure developers from traditional programming backgrounds can leverage their existing expertise while accessing the full capabilities of blockchain environments with minimal friction.

\subsection{Security Considerations}
\label{sec:security_considerations}
Security is paramount in blockchain systems and smart contract execution environments. DTVM employs a multi-layered security design, leveraging WebAssembly’s inherent sandboxing capabilities and integrating specific safeguards into its JIT compiler and runtime. This section outlines DTVM’s security measures, contrasts its approach with alternative methodologies, and details its implementation.  

\textbf{Trade-offs Between AOT and JIT Compilation}

While ahead-of-time (AOT) compilation has been explored for performance gains by precompiling contracts, it poses challenges for general-purpose smart contract platforms. Key limitations include:  
\textit{Trust and Validation Complexity}: Ensuring all nodes use identical, trusted compiler versions and configurations requires complex multi-party authentication or reliance on centralized trusted compilers, introducing additional trust assumptions and attack vectors.  
\textit{Limited Universality}: AOT optimizations are typically restricted to whitelisted contracts (e.g., system contracts), failing to address performance needs for arbitrary user-deployed contracts on permissionless blockchains.

In contrast, just-in-time (JIT) compilation—used by DTVM and similar systems—balances performance and flexibility, supporting arbitrary contract execution. However, JIT introduces its own risks, such as enlarged attack surfaces due to dynamic code generation, compiler vulnerabilities, and potential code-injection risks from improper memory/execution management.

\textbf{DTVM’s Security Mechanisms}

DTVM mitigates these risks through the following foundational safeguards:  
\begin{itemize}  
  \item \textbf{Constrained Control Flow}: JIT-generated machine code adheres to strict control flow rules. Function addresses and jump/call targets are tightly controlled. Transfers are limited to within compiled code blocks or predefined, VM Native Interface (VNI)-managed host API entry points. No mechanisms allow jumps to arbitrary memory addresses provided by contracts or external inputs, preventing Arbitrary Code Execution (ACE) attacks.
  \item \textbf{Memory Execution Confinement}: JIT-compiled code executes in mmap-allocated memory regions with \lstinline|PROT_READ | | \lstinline|PROT_EXEC| permissions (write permissions are restricted to brief, strictly managed code-patching phases). This executable memory is isolated from WebAssembly linear memory used for contract data. Hardware-enforced boundary checks (e.g., guard pages described in Section 4.3) further restrict data access. Unauthorized execution outside designated regions or data access beyond valid memory boundaries triggers hardware exceptions (e.g., SIGSEGV), safely handled by the VM runtime to produce deterministic traps rather than security breaches.
  \item \textbf{Deterministic Execution Foundation (dWasm)}: As detailed in Section 4.6, dWasm eliminates uncertainty sources (e.g., ambiguous floating-point NaN representations or uncontrolled host function interactions). Strict validation and resource limits applied before JIT compilation further reduce attack surfaces.
  \item \textbf{Integrated Safety Checks}: Critical security checks (e.g., integer overflow detection from Section 4.5 and gas metering from Section 4.4) are embedded directly into JIT compilation. This ensures robust enforcement of foundational safety properties, mitigating common vulnerabilities like overflow exploits or denial-of-service attacks via resource exhaustion.
\end{itemize}

\textbf{Attack Mitigation and Network-Level Security}

Even if an attacker compromises a single validator node (e.g., via OS-level vulnerabilities or undetected JIT flaws), the impact is limited.  The compromised node’s incorrect state transitions or invalid transaction validations would conflict with honest majority nodes during consensus checks. Faulty nodes are likely excluded or penalized by consensus protocols, preserving the blockchain’s overall integrity. This design ensures DTVM’s execution environment is secure at both the node level and the network consensus layer, balancing performance, flexibility, and safety.  

\subsection{AI-Assisted Development Tools}
DTVM Stack integrates a full-stack suite of AI-driven tools to support the entire lifecycle of smart contract development, from coding and debugging to security auditing, deployment, and invocation. 
This includes MCP services that can be independently deployed and integrated with open-source large language models (LLMs). Extensive efforts have been dedicated to data preprocessing, model fine-tuning, and RAG methodologies to enhance functionality. 
Specific optimizations on the Qwen model ~\cite{bai2023qwen} significantly improve code generation efficiency, security audit capabilities and code dependencies analysis accuracy.  

\textbf{Development Stage Enhancements}

To accelerate smart contract development, DTVM provides AI-assisted code generation and compilation/deployment tooling. We have built a high-quality contract code corpus to mitigate LLM hallucination-induced code chaos, such as Solidity syntax version inconsistencies and tight project dependency coupling. Our solution supports generating complex smart contracts for auctions, gaming, DeFi, and other domains.

\textbf{Security Auditing with Multi-Agent Workflows}

A multi-agent collaborative architecture is employed for security audits, integrating optimized audit prompts, strategies, and vulnerability databases to significantly enhance detection efficiency. Comparative evaluations against baseline models and existing tools demonstrate substantial improvements in performance. To establish a baseline, we directly used API calls to Qwen-Max without employing specialized workflows. Additionally, GPTLens~\cite{hu2023large}, a well-known open-source tool with audit-only functionality, was included in the evaluation for comparison. The results underscore the superiority of DTVM’s integrated multi-agent approach, both in terms of vulnerability detection and automated resolution capabilities. 

\textbf{Automated Code Repair Capabilities}

For code correction, DTVM’s tools achieve an 86\% repair success rate, surpassing both baseline models and specialized tools. As shown in Table \ref{table:security_aduit_perf}, we used Qwen-Plus mode as the evaluation baseline for code repair and compared its performance to SolGPT, an open-source tool specifically designed for Solidity code correction.The improvements achieved by DTVM can be attributed to several key factors. Tailored Prompt Engineering: Customized prompts that effectively capture the nuances of repair requirements.
Domain-Specific Knowledge Integration: Incorporation of expertise specific to smart contract development, enhancing the accuracy and relevance of repairs. Iterative Refinement of Repair Strategies: Continuous optimization of repair strategies based on feedback, leading to incremental improvements in success rates.These advancements enable DTVM to outperform existing baseline models and demonstrate superior capabilities compared to other specialized tools. 

\textbf{Accurate Code Dependencies Analysis}

A specialized agent is provided for code compilation, enabling accurate extraction of the dependencies imported in the target contract file. We established a knowledge base by fetching the release information of mainstream dependencies from GitHub offline, segmenting it into binary blocks, and performing embedding. By integrating prompt templates with the dedicated knowledge base, RAG is achieved, increasing the accuracy of dependencies extraction to over 95\%. In contrast, when relying solely on prompts, the accuracy is only 65\% with Qwen-Plus and 75\% with Qwen-Max.

\begin{table}
  \centering
  \caption{Security Audit Performance Comparison}
  \begin{tabular}{|c|c|c|c|}
  \hline
 \textbf{Metric} & \textbf{Our Tool} & \textbf{Baseline (Qwen-Max 20250407)} & \textbf{GPTLens (Open-source Audit Tool)} \\
  \hline
  \textbf{Detection Rate} & $81\%$ & $5\%$ & $24\%$  \\
  \hline 
  \textbf{False Positive Rate} & $27\%$ & $82\%$ & $73\%$ \\
  \hline
  \end{tabular}
  \label{table:security_aduit_perf}
\end{table}

\begin{table}
  \centering
  \caption{Code Repair Performance Comparison}
  \begin{tabular}{|c|c|}
  \hline
 \textbf{Tool} & \textbf{Repair Success Rate} \\
  \hline
  \textbf{Our Tool} & $86\%$ \\ 
  \hline
  \textbf{Baseline (Qwen-Plus 20250407)} & $71\%$ \\
  \hline
  \textbf{SolGPT} & $43\%$ \\
  \hline
  \end{tabular}
  \label{table:security_aduit_perf}
\end{table}

\section{Evaluation}\label{sec: Evaluation}
We evaluate the performance of \tool by benchmarking it against current mainstream EVM-based and Wasm-based VMs. Our evaluation mainly addresses the following key aspects: (1) functional validation, and (2) performance improvement of \tool compared to other VMs in various scenarios.

\paragraph{Baselines} We compare \tool with several competitive baseline VMs, including interpreted‌-based evmone\footnote{https://github.com/ethereum/evmone}, JIT EVM Revmc\footnote{https://github.com/paradigmxyz/Revmc}, and JIT Wasm VMs: Wasmtime\footnote{https://github.com/bytecodealliance/wasmtime} and Wasmer\footnote{https://github.com/wasmerio/wasmer}.

\begin{itemize}
    \item[$\bullet$] Evmone (v0.13.0): Evmone is a high-performance, open-source C++ implementation of the Ethereum Virtual Machine. Given its  high efficiency and full compatibility with the Ethereum protocol, evmone is widely used in various Ethereum clients for executing Ethereum-style smart contracts.

    \item[$\bullet$] Revmc (v0.1.0): Revmc implements experimental JIT\&AOT compilers for EVM to accelerate execution by lowering the EVM bytecode. It provides native integration with the Revm virtual machine. Revmc proposes its optimized intermediate representation and leverages LLVM as its compiler backend for code generation. Given that Revmc is currently at experimental phase,  the comparison relies solely on its provided data for a general evaluation.


    
    
    \item[$\bullet$] Wasmtime (v31.0.0): Wasmtime, developed by the Bytecode Alliance, serves as a Rust-written Wasm VM , providing a lightweight Wasm execution environment. Specifically, Wasmtime features a custom JIT compiler using the Cranelift code generator, which further optimizes execution efficiency. In our evaluation, we compare \tool with Wasmtime utilizing Cranelift JIT compiler (referred to as Wasmtime (Cranelift)).
    
    \item[$\bullet$] Wasmer (v5.0.5): Wasmer offers an efficient and secure Wasm VM, which is written in Rust and already open-sourced. Notably, Wasmer integrates different compiler backends for optimized Wasm execution: (1) Wasmer (Singlepass), which offers custom JIT compilation; (2) Wasmer (Cranelift), backed by Cranelift; and (3) Wasmer (LLVM), which employs the LLVM compiler infrastructure. We conducted comparisons of \tool against these three Wasmer variants during the evaluation.

\end{itemize}


\begin{table}[h]
    \centering
    \renewcommand{\arraystretch}{1.3} 
    \caption{Benchmarking workloads \& categories}
    \vspace{5pt}
    \begin{tabular}{|c|l|c|}
        \hline
        & \textbf{Workload} & \textbf{Category} \\ 
        \hline
        \textbf{A1} & ERC20 & \multirow{5}{*}{Semantic-Rich Smart Contracts} \\ 
        \cline{1-2}
        \textbf{A2} & ERC721& \\ 
        \cline{1-2}
        \textbf{A3} & ERC1155& \\ 
        \cline{1-2}
        \textbf{A4} & Counter& \\ 
        \cline{1-2}
        \textbf{A5} & Integer Overflow & \\ 
        \hline
        \textbf{B1} & Fibonacci & \multirow{1}{*}{Compute-Intensive Smart Contracts} \\ 
        \hline 
        \textbf{C1} & PolyBench & \multirow{2}{*}{Standard Benchmark Suits} \\ 
        \cline{1-2}
        \textbf{C2} & WAPM & \\ 
        \hline 
    \end{tabular}
    \label{tab: benchmark}
\end{table}


\paragraph{Benchmarks} To fully evaluate the advantages of \tool, we selects diverse workloads and divide them into three categories: \emph{Workload A} includes various semantic-rich smart contracts, \emph{Workload B} comprises several compute-intensive smart contracts, and \emph{Workload C} prepares two standard benchmark suits, as depicted in Table~\ref{tab: benchmark}.

\begin{itemize}
    \item[$\bullet$] \textbf{Workload A: Semantic-Rich Smart Contracts.} This category is used to evaluate common on-chain smart contract workloads, which consists of five popular smart contracts: ERC20, ERC721, ERC1155, Counter, and Integer Overflow. More precisely, the ERC20 contract implements token transfer (TR) logic. ERC721 and ERC1155 establish standards for transferring non-fungible tokens, using functions such as transferFrom and safeBatchTransferFrom (collectively referred to as NFT-TRF). For Counter contract, its increase function (INC) reads data from the host, applies simple math, and updates the results to the host. And the Integer Overflow contract constructs an arithmetic overflow scenario by constructing a Uniswap-like token swap case.
    

    \item[$\bullet$] \textbf{Workload B: Compute-Intensive Smart Contracts.} This category includes a typical compute-intensive smart contract example: the Fibonacci contract, whose function (FIB) recursively calculates the Fibonacci sequence for a given input number.
    

    \item[$\bullet$] \textbf{Workload C: Standard Benchmark Suits. } This category incorporates two standard benchmark suites: PolyBench (v4.2.1)\footnote{https://github.com/MatthiasJReisinger/PolyBenchC-4.2.1} and WAPM\footnote{https://github.com/wapm-packages}, aiming to measure VM's runtime performance under diverse computational scenarios. Specifically, PolyBench quantifies the computational efficiency when processing numerically intensive tasks, such as matrix operations and linear algebra. We use the PolyBenchC implementation and compile it into Wasm modules to conduct comparative tests. WAPM provides various useful Wasm packages. From these, we select a few typical packages (see Appendix~\ref{appx: wapm}) to assess and compare the latency to first invocation of baseline VMs.
\end{itemize}


\paragraph{Testbed and Setup} Our experiments are conducted on virtual machines, each equipped with 2.70GHz Intel(R) Xeon(R) Platinum 8369B CPU processor (8 vCPUs, 16 GiB RAM). The machines run CentOS 7.9 with 3.10.0 Linux Kernel. In the evaluation of semantic-rich contracts, we leverage OpenZeppelin's open-source implementations of the ERC 20/721/1155, written in Solidity. We utilize Solc (v0.8.25) with optimizer option enabled to compile these contracts into EVM bytecode for execution via evmone. Additionally, we use \tool's built-in Solidity-to-Wasm generator to convert the EVM bytecode into Wasm bytecode for execution on \tool. For compute-intensive ones, we provide corresponding source code written in Solidity/C++, compile it into Wasm bytecode, and execute it using baseline VMs. We implement mocked host APIs purely in-memory to avoid storage impact on the results.

\paragraph{Measure Metrics} For functional validation, we compare \tool and other VMs based on the following metrics: (1) Smart contract language support, (2) Determinism guarantee, (3) Machine code compactness, and (4) Binary artifact size. 

For performance evaluation, we measure the following metrics: (1) Latency to first invocation (post-loading \& compilation): the time from loading the Wasm module binary until it is ready to be invoked, and (2) Execution latency: the duration taken by the VM to execute smart contracts/functions, and (3) Processing time: the total time comprising latency to first invocation and execution latency, i.e., the sum of metrics (1) and (2).

\begin{table}[t]
\centering
\renewcommand{\arraystretch}{1.3}
\caption{Overview of Smart Contract Language Support} \label{tab: language support}
\vspace{5pt}
\begin{tabularx}{\textwidth}{XXXXXXp{2.5cm}}
 \toprule
 \makecell[c]{\textbf{Baselines}}
 & \makecell[c]{\textbf{Solidity}} 
 & \makecell[c]{\textbf{C/C++}} 
 & \makecell[c]{\textbf{Java}} 
 & \makecell[c]{\textbf{Rust}} 
 & \makecell[c]{\textbf{Golang}}
 & \makecell[c]{\textbf{AssemblyScript}}\\
    \midrule

     \makecell[l]{\textbf{Evmone}} & \makecell[c]{\Checkmark} & \makecell[c]{\XSolidBrush} & \makecell[c]{\XSolidBrush} & \makecell[c]{\XSolidBrush}& \makecell[c]{\XSolidBrush} & \makecell[c]{\XSolidBrush}\\

    \makecell[l]{\textbf{Revmc}} & \makecell[c]{\Checkmark} & \makecell[c]{\XSolidBrush} & \makecell[c]{\XSolidBrush} & \makecell[c]{\XSolidBrush}& \makecell[c]{\XSolidBrush} & \makecell[c]{\XSolidBrush}\\
      
     \makecell[l]{\textbf{Wasmtime}} & \makecell[c]{\XSolidBrush} & \makecell[c]{\Checkmark} & \makecell[c]{\Checkmark\kern-1.2ex\raisebox{1ex}{\rotatebox[origin=c]{125}{\textbf{--}}}} & \makecell[c]{\Checkmark}& \makecell[c]{\Checkmark} & \makecell[c]{\Checkmark\kern-1.2ex\raisebox{1ex}{\rotatebox[origin=c]{125}{\textbf{--}}}}\\
     
     \makecell[l]{\textbf{Wasmer}} & \makecell[c]{\XSolidBrush} & \makecell[c]{\Checkmark} & \makecell[c]{\Checkmark\kern-1.2ex\raisebox{1ex}{\rotatebox[origin=c]{125}{\textbf{--}}}} & \makecell[c]{\Checkmark}& \makecell[c]{\Checkmark} & \makecell[c]{\Checkmark\kern-1.2ex\raisebox{1ex}{\rotatebox[origin=c]{125}{\textbf{--}}}}\\

    \makecell[l]{\textbf{\tool}} & \makecell[c]{\Checkmark} & \makecell[c]{\Checkmark} & \makecell[c]{\Checkmark} & \makecell[c]{\Checkmark}& \makecell[c]{\Checkmark} & \makecell[c]{\Checkmark}\\
     
    \bottomrule
\end{tabularx}
\footnotesize

\begin{itemize}
    \item[{\centering[\Checkmark]}]  Full support for the smart contract language.
    \item[{\centering[\XSolidBrush]}] No support for the smart contract language.
    \item[{\centering[\Checkmark\kern-1.2ex\raisebox{1ex}{\rotatebox[origin=c]{125}{\textbf{--}}}]}] Support for the general-purpose language, but lacks smart contract-specific features. 
\end{itemize}

\end{table}

\begin{table}[h]
\centering
\small
\renewcommand{\arraystretch}{1.3}
\caption{Comparison of deterministic execution}
\label{tab: Comparison of deterministic execution}
\vspace{5pt}
\begin{tabular}{p{3cm}p{6.1cm}p{6.1cm}
    >{\raggedright}p{2.8cm}  
    >{\raggedright}p{6cm}    
    >{\raggedright}p{4.2cm}  
} 
    \toprule
    \textbf{Baselines} & \makecell[c]{\textbf{Intel X86-64}} & \makecell[c]{\textbf{ARM}} \\
    \midrule

    \textbf{Wasmtime (Cranelift)} 
    &
    
    \begin{minipage}[t]{\linewidth}
    \vspace{-1.3em} 
      \begin{lstlisting}[numbers=none]
// Wasm backtrace 
0: <wasm function 0>
1: 0x3b <wasm function 0>
...
!!32718: 0x47 <wasm function 1>!!
wasm trap: call stack exhausted
    \end{lstlisting}
    \end{minipage}
    &
    
    \begin{minipage}[t]{\linewidth}
    \vspace{-1.3em} 
      \begin{lstlisting}[numbers=none]
// Wasm backtrace 
0: <wasm function 0>
1: 0x3b <wasm function 0>
...
32718: 0x3b <wasm function 0>
32719: 0x3b <wasm function 0>
32720: 0x3b <wasm function 0>
@@32721: 0x47 <wasm function 1> @@
wasm trap: call stack exhausted
      \end{lstlisting}
    \end{minipage} 
    \\
    \midrule
    \textbf{Wasmer (Singlepass)} 
    &
    
    \begin{minipage}[t]{\linewidth}
    \vspace{-1.3em} 
      \begin{lstlisting}[numbers=none]
// Wasm backtrace 
...
!!backtrace depth: 13098!!
RuntimeError: call stack exhausted
    \end{lstlisting}
    \end{minipage}
    &
    
    \begin{minipage}[t]{\linewidth}
    \vspace{-1.3em} 
      \begin{lstlisting}[numbers=none]
// Wasm backtrace 
...
@@backtrace depth: 10926@@
RuntimeError: call stack exhausted
      \end{lstlisting}
    \end{minipage} 
    \\
    \midrule
    \textbf{Wasmer (Cranelift)}
    &    
    \begin{minipage}[t]{\linewidth}
    \vspace{-1.3em} 
      \begin{lstlisting}[numbers=none]
// Wasm backtrace 
...
!!backtrace depth: 65462!!
RuntimeError: call stack exhausted
    \end{lstlisting}
    \end{minipage}
    &
    
    \begin{minipage}[t]{\linewidth}
    \vspace{-1.3em} 
      \begin{lstlisting}[numbers=none]
// Wasm backtrace 
...
@@backtrace depth: 65530@@
RuntimeError: call stack exhausted
      \end{lstlisting}
    \end{minipage} 
    \\

    \midrule
    \textbf{\tool} 
    &
    
    \begin{minipage}[t]{\linewidth}
    \vspace{-1.3em} 
      \begin{lstlisting}[numbers=none]
// Wasm backtrace 
#00  $f00
#01  $f00
...
<<#15  $f00<<
error_msg: WasmCallStackExceed
      \end{lstlisting}
    \end{minipage}
    &
    
    \begin{minipage}[t]{\linewidth}
    \vspace{-1.3em} 
      \begin{lstlisting}[numbers=none]
// Wasm backtrace 
#00  $f00
#01  $f00
...
<<#15  $f00<<
error_msg: WasmCallStackExceed
      \end{lstlisting}
    \end{minipage} 
    \\
    
    \bottomrule
\end{tabular}
\footnotesize
\begin{itemize}
    \item[$\bullet$] Wasmtime exhibits different stack overflow depths across Intel X86-64 and ARM architectures. Specifically, the stack exceeded at call depth 32718 on Intel X86-64, while the depth 32721 on ARM.

    \item[$\bullet$] Wasmer variants also produce different stack overflow depths across two architectures: Wasmer (Singlepass) cores at depth of 13098 and 10926, while Wasmer (Cranelift) backtraces at separate call depth 65462 and 65530.

    \item[$\bullet$] \tool maintains consistent stack overflow behaviors across both architectures.
\end{itemize}
\vspace{-16pt}
\end{table}

\subsection{Functional Validation}\label{subsec: Functional Validation}

\paragraph{Smart Contract Language Support} We evaluate the smart contract languages supported by \tool, evmone, revmc, Wasmer, and Wasmtime. As shown in Table~\ref{tab: language support}, evmone and revmc are specifically tailored to support Solidity, the primary language for Ethereum smart contracts. However, Wasmtime and Wasmer lack Solidity-to-Wasm compilation and execution capabilities, which limits their use in Ethereum ecosystems. Nevertheless, both Wasmtime and Wasmer support smart contracts written in C/C++, Rust, and Golang, making them suitable for non-Solidity environments. Additionally, while they provide general-purpose Java/AssemblyScript support, they do not fully adapt to the specific requirements and features of corresponding smart contracts. In contrast, \tool supports a wide spectrum of smart contract languages, including C/C++, Java, Rust, Golang, and AssemblyScript, and seamlessly provides robust compatibility with the Solidity-based Ethereum ecosystem via its optimized Solidity-to-Wasm generator and adaptive runtimes. The broad language support of \tool validates its potential to provide a unified development ecosystem for smart contracts.

\paragraph{Determinism Guarantee} We evaluate the deterministic execution guarantee of \tool across the Intel X86-64 and ARM CPU architecture, contrasting it with Wasmtime and Wasmer variants. To assess determinism, we construct an edge-case scenario (infinite loop) that might exhibit non-deterministic behaviors across different CPU architecture. 

As depicted in Table~\ref{tab: Comparison of deterministic execution}, Wasmtime and Wasmer variants exhibit inconsistent stack overflow depths, violating the deterministic execution guarantee required by blockchain systems. Such non-determinism arises due to differences in CPU architectures, including variations in register design and default stack sizes. These factors cause the actual stack consumption of the same Wasm function to differ, leading to unpredictable stack overflow boundaries. On the contrary, \tool demonstrates consistent behaviors across Intel X86-64 and ARM through its proposed deterministic mechanisms, making it ideal for blockchain scenarios.

\paragraph{Machine Code Compactness} We compare the size of the machine code generated by different compiler backends to showcase the compactness advantage of \tool. In this comparison, we compile Fibonacci (B1) and selected cases from PolyBench (C1) to assess both the object size (obj) and the code size (code) of the resulting machine code. The object size measures the total size of the intermediate target file, encompassing the code segment, data segment, symbol table, and relocation table, representing the storage size and loading time of the target file. This metric impacts both the storage requirements and the loading time. The code size denotes the actual size of the machine code, which represents the code generation efficiency. 

\begin{table}[h]
    \centering
    \small
    \renewcommand{\arraystretch}{1.3}
    \caption{Comparison of compiled machine code size (in KBytes)} \label{tab: Comparison of machine code size}
    \vspace{5pt}
    \begin{tabularx}{\linewidth}{@{} >{\raggedright\arraybackslash}X *{10}{r} @{}}
    \toprule
    \multirow{2}{*}{\textbf{Baselines}} & \multicolumn{2}{c}{\textbf{Fibonacci}} & \multicolumn{2}{c}{\textbf{Jacobi-2d}} & \multicolumn{2}{c}{\textbf{2mm}} & \multicolumn{2}{c}{\textbf{Adi}} & \multicolumn{2}{c}{\textbf{Symm}} \\
    \cmidrule(lr){2-3} \cmidrule(lr){4-5} \cmidrule(lr){6-7} \cmidrule(lr){8-9} \cmidrule(lr){10-11}
    & \makecell[c]{obj} & \makecell[c]{code} & \makecell[c]{obj} & \makecell[c]{code} & \makecell[c]{obj} & \makecell[c]{code} & \makecell[c]{obj} & \makecell[c]{code} & \makecell[c]{obj} & \makecell[c]{code} \\
    \midrule
   
    \makecell[l]{\textbf{Wasmtime (Cranelift)}} & 13.3 & 0.3 & 123.4 & 48.0 & 123.4 & 48.5 & 127.4 & 49.5 & 123.4 & 48.3 \\
    \makecell[l]{\textbf{Wasmer (Singlepass)}} & 4.4 & 0.8 & 256.6 & 126.1 & 259.0 & 127.4 & 266.8 & 130.5 & 257.8 & 127.0 \\
    \makecell[l]{\textbf{Wasmer (Cranelift)}} & 4.1 & 0.3 & 137.0 & 45.6 & 138.2 & 46.0 & 140.2 & 47.1 & 137.7 & 45.9 \\
    \makecell[l]{\textbf{Wasmer (LLVM)}} & 3.5 & 0.2 & 65.6 & \textbf{34.7} & 66.0 & \textbf{35.0} & 67.0 & \textbf{35.9} & 65.7 & \textbf{34.8} \\
    \makecell[l]{\textbf{\tool}} & \textbf{1.0} & \textbf{0.2} & \textbf{57.2} & 46.2 & \textbf{49.2} & 38.5 & \textbf{50.1} & 39.7 & \textbf{49.2} & 38.5 \\
    
    \bottomrule
    \end{tabularx}
    \footnotesize
\end{table}

As demonstrated in Table~\ref{tab: Comparison of machine code size}, \tool incurs the lowest object size in all tested scenarios, this highlights that \tool has the minimized storage requirement and the fastest loading time for the target file among all baselines. In terms of code size measurement, \tool's output size is relative smaller compared to Wasmtime (Cranelift), Wasmer (Cranelift) and Wasmer (Singlepass), saves up to 30.9\%, 30.0\%, and 72.6\%, respectively. Notably, while the code size of \tool is slightly larger than that of Wasmer (LLVM) in some scenarios, it remains significantly smaller than that of other compiler backends. This indicates that \tool effectively reduces instruction redundancy, despite omitting some optimizations compared to LLVM.

\begin{table}[h]
\centering
\small
\renewcommand{\arraystretch}{1.3}
\caption{Comparison of codebase and artifact size} \label{tab: Comparison of artifact size}
\vspace{5pt}
\begin{tabularx}{\textwidth}{p{3cm}XX}
    \toprule
   \makecell[l]{\textbf{Baselines}}
 & \makecell[c]{\textbf{Codebase LoC (K lines)}} 
 & \makecell[c]{\textbf{Stand-Alone CLI Size (MB)}} \\
     \midrule
     \makecell[l]{\textbf{Wasmtime (Cranelift)}} & \makecell[c]{143.6} & \makecell[c]{47.6}\\
     \makecell[l]{\textbf{Wasmer (Singlepass)}} & \makecell[c]{314.8} & \makecell[c]{48.1}\\
     \makecell[l]{\textbf{Wasmer (Cranelift)}} & \makecell[c]{289.0} & \makecell[c]{50.9}\\
     \makecell[l]{\textbf{Wasmer (LLVM)}} & \makecell[c]{299.0} & \makecell[c]{111.5}\\
     \makecell[l]{\textbf{\tool}} & \makecell[c]{69.5} & \makecell[c]{31.8}\\
    \bottomrule
\end{tabularx}
\end{table}

\paragraph{Lightweight Artifacts} We analyze the codebase's lines of code (LoC) and the size of the client binaries to demonstrate \tool's advantages in lightweight deployment. As observed in Table~\ref{tab: Comparison of artifact size}, the core implementation of \tool comprises 69.5 KLoC, which remains dramatically smaller than the codebases of Wasmtime (Cranelift) (143.6 KLoC), Wasmer (Singlepass) (314.8 KLoC), Wasmer (Cranelift) (289.0 KLOC), and Wasmer (LLVM) (299.0 KLoC). Note that, the minimized codebase ensures the smallest Trusted Computing Base (TCB) when integrated \tool into TEEs, thereby enhancing the security and efficiency. Also, \tool's 31.8MB CLI artifact size surpasses all baselines in compactness, achieving 33.1\% $\sim$ 71.6\% reduction over other baselines. This further validates \tool’s superior efficiency in resource-constrained lightweight deployment.




\begin{figure}[h]
    \setlength{\abovecaptionskip}{0cm}
    \setlength{\belowcaptionskip}{-1.5\baselineskip}
    \centering
    \subfigure[ERC20]{\label{fig: exp1.2.1a}
    \includegraphics[width=0.4\columnwidth]{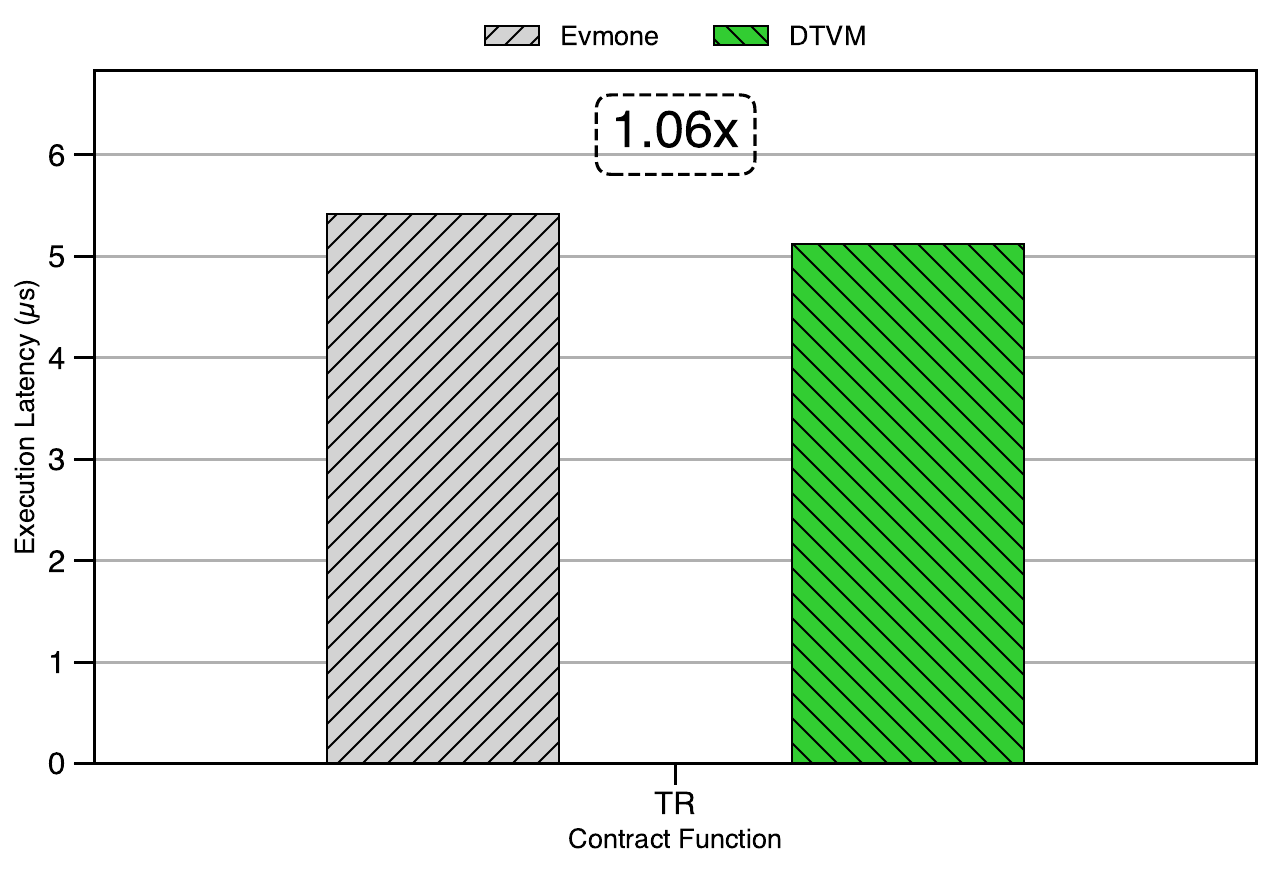}}
    \hspace{0.05in} 
    \subfigure[ERC721]{\label{fig: exp1.2.1b}
    \includegraphics[width=0.4\columnwidth]{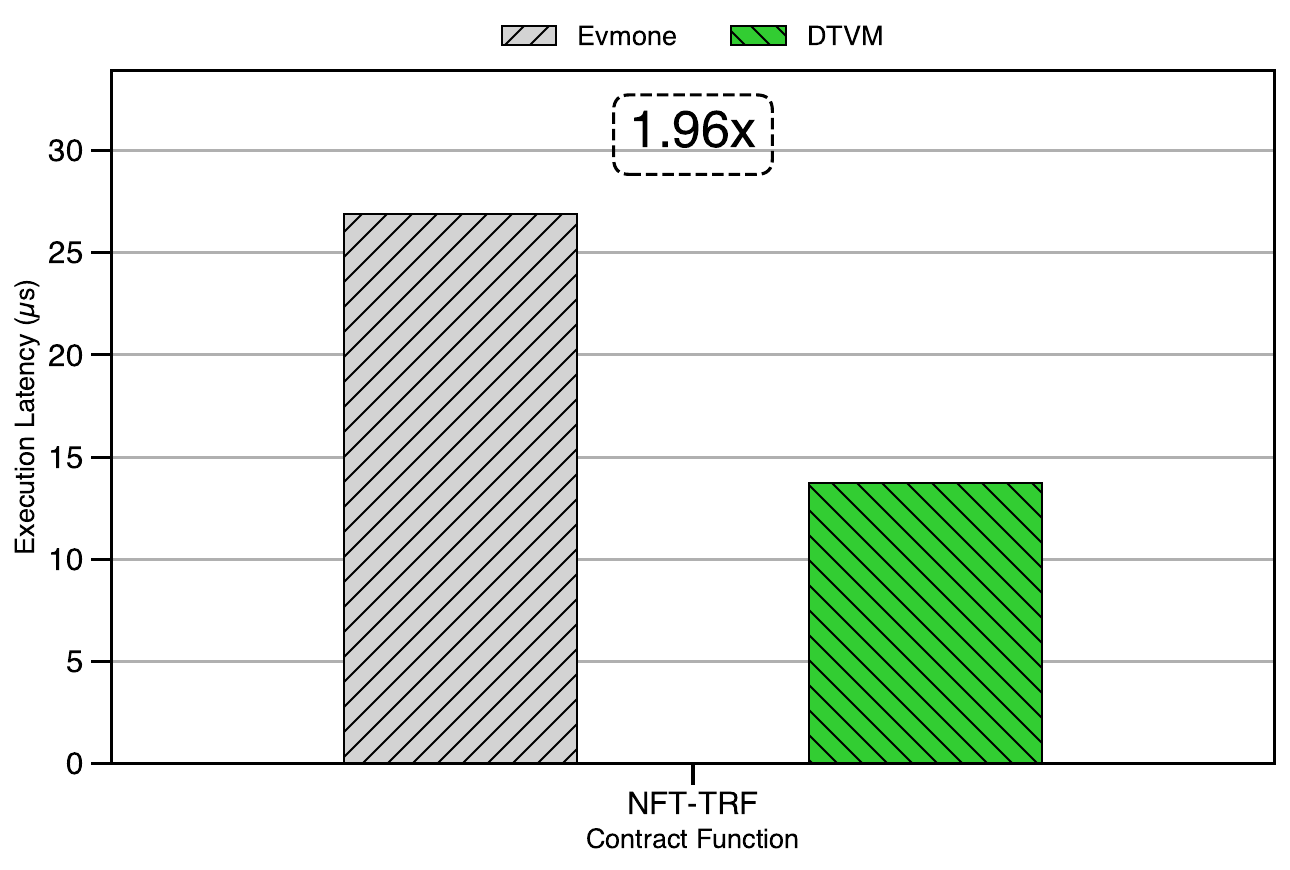}}
    \subfigure[ERC1155]{\label{fig: exp1.2.1c}
    \includegraphics[width=0.4\columnwidth]{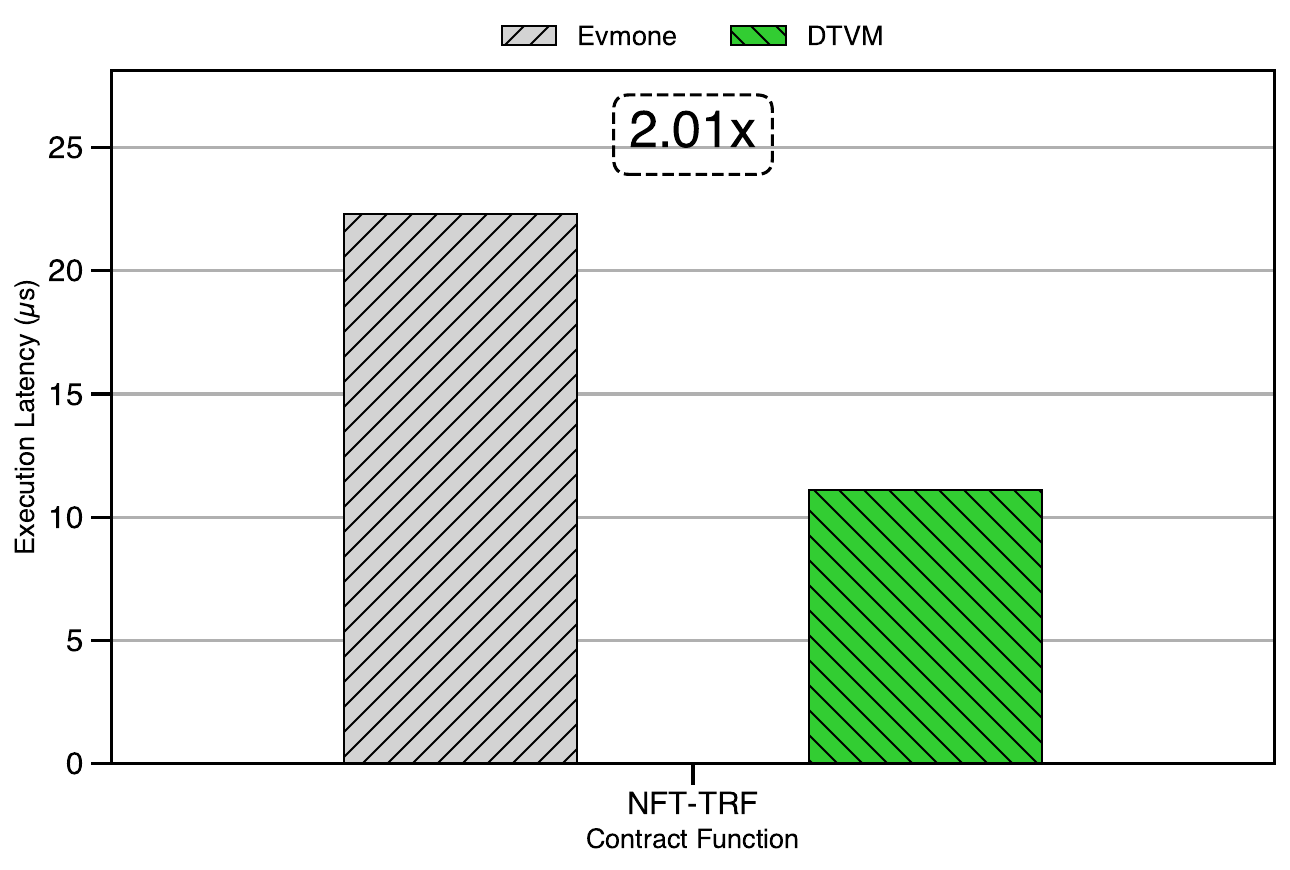}} 
    \subfigure[Fibonacci(25)]{\label{fig: exp1.2.1e}
    \includegraphics[width=0.4\columnwidth]{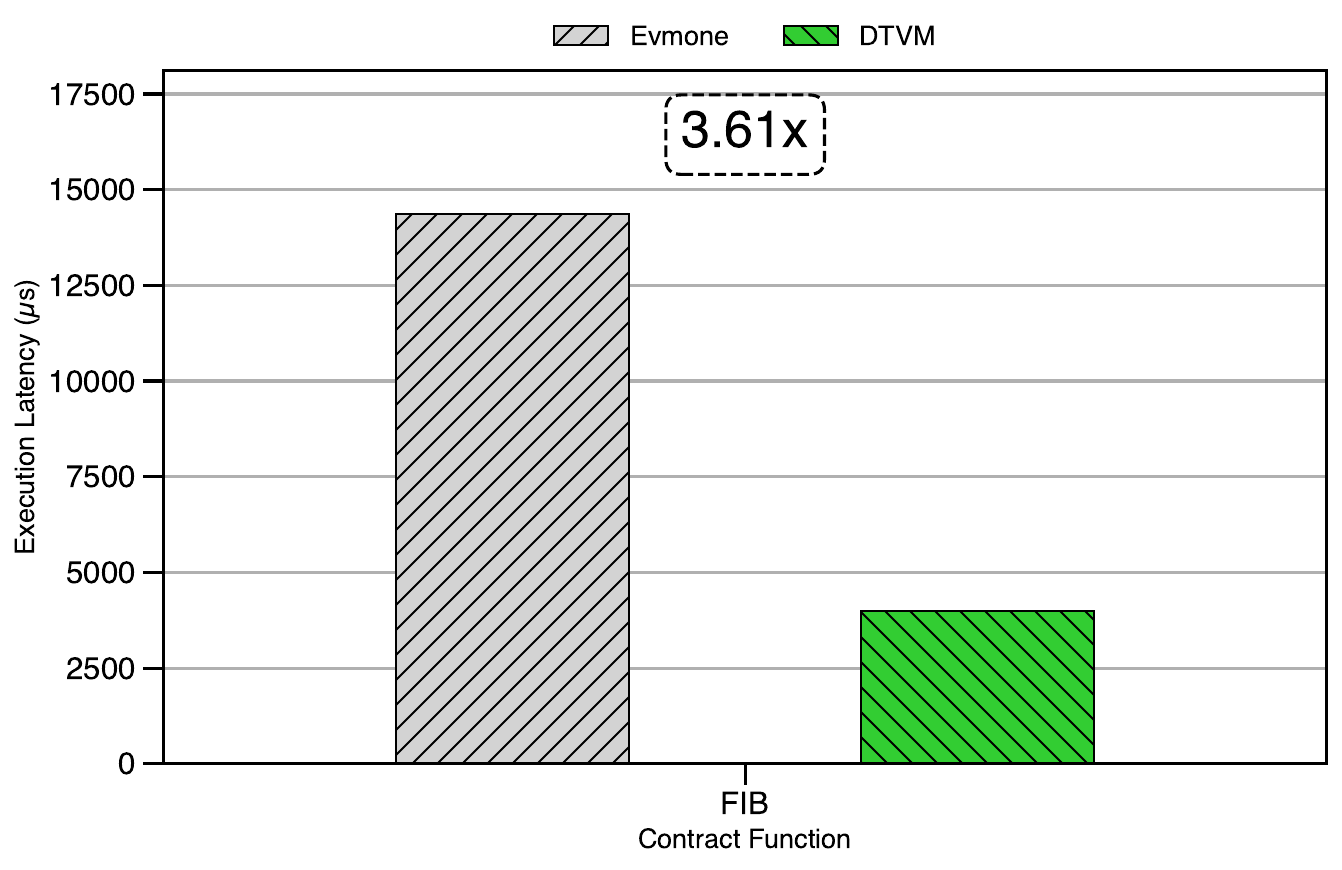}}
    \caption{Comparison in the Ethereum Ecosystem} \label{fig: exp1.2.1}
\end{figure}

\subsection{Performance Evaluation}\label{subsec: Performance Evaluation}

\subsubsection{Comparison in the Ethereum ecosystem} 
We utilize cases from Workload A and Workload B to measure execution latency, comparing the performance of DTVM with e and Revmc to realize a comparison within the Ethereum ecosystem. For each experiment, we perform 10,000 serial executions and report the average time taken for each smart contract execution. The related experimental results are shown in the Fig.~\ref{fig: exp1.2.1}.

\paragraph{\tool in Comparison with Evmone} For token transfers in ERC20, as shown in Fig.~\ref{fig: exp1.2.1a}, \tool incurs the lowest execution latency with 5.12µs, resulting in 6\% latency cut down compared to evmone (5.42$\mu$s). The results in Fig.~\ref{fig: exp1.2.1b} and Fig.~\ref{fig: exp1.2.1c} show that \tool achieves 1.96$\times$ and 2.01$\times$ speedups over evmone in executing ERC721 and ERC1155 NFT transfers, respectively. Regarding compute-intensive contracts, in the Fibonacci workload, \tool also incurs a notable 3.61$\times$ speedup over evmone (see Fig.~\ref{fig: exp1.2.1e}). These results further validate that \tool remains superior efficiency, and utilizing \tool can significantly accelerate smart contract execution. 

\begin{figure}[h]
    \setlength{\abovecaptionskip}{0cm}
    \setlength{\belowcaptionskip}{-1.5\baselineskip}
    \centering
    \includegraphics[width=0.7\columnwidth]{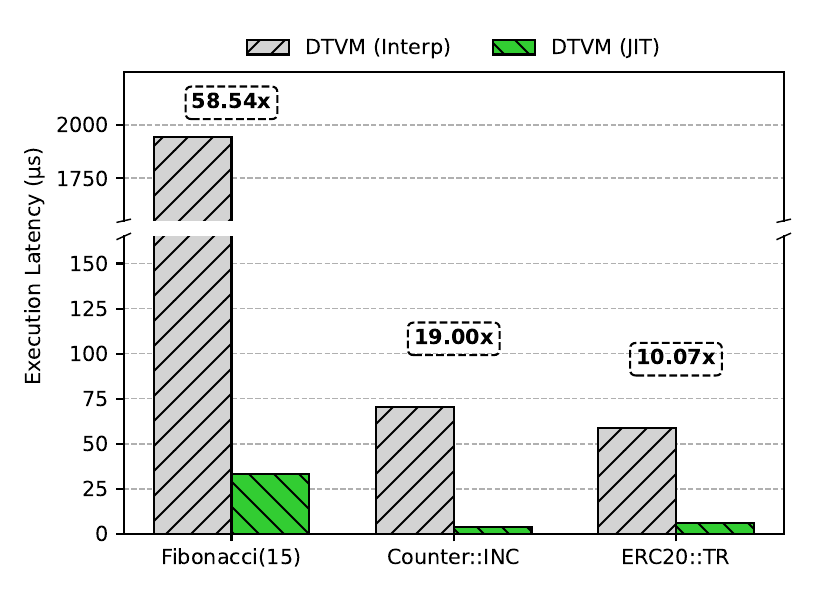}
    \caption{\tool(JIT) improvement over \tool(Interp)} \label{fig: exp1.2.1f}
    \vspace{16pt}
\end{figure}

\paragraph{JIT Compilation vs. Interpretation} We also compare the execution efficiency of \tool using JIT compilation, denoted as \tool (JIT), in comparison with interpretation, denoted as \tool (Interp). The experimental results are presented in Fig.~\ref{fig: exp1.2.1f}. As can be observed, \tool (JIT) exhibits a 58.54$\times$ performance improvement in executing the Fibonacci contract with input 25. This highlights the benefits of \tool's JIT compilation, which generates optimized machine code to accelerate compute-intensive workloads. Besides, \tool (JIT) also gains notable performance improvements for the Counter (19.00$\times$) and ERC20 (10.07$\times$) smart contracts. This further demonstrates the high efficiency of \tool's JIT compiler. Additionally, regarding the acceleration factor of JIT relative to interpretation, we compare \tool with Remvc (using their reported performance data\footnote{https://github.com/paradigmxyz/revmc/assets/17802178/96adf64b-8513-469d-925d-4f8d902e4e0a}) and found that \tool's JIT achieves a higher acceleration factor than that of Remvc.

\subsubsection{Comparison with General-Purpose Wasm VMs} 
In this evaluation, we compare \tool that disables the lazy compilation option with current SOTA Wasm VMs by measuring processing time which includes the compilation and execution time.

\begin{figure}[ht]
    \setlength{\abovecaptionskip}{0cm}
    \setlength{\belowcaptionskip}{-1.5\baselineskip}
    \centering
    \subfigure[Fibonacci(30)]{\label{fig: exp1.2.2a}
    \includegraphics[width=0.4\columnwidth]{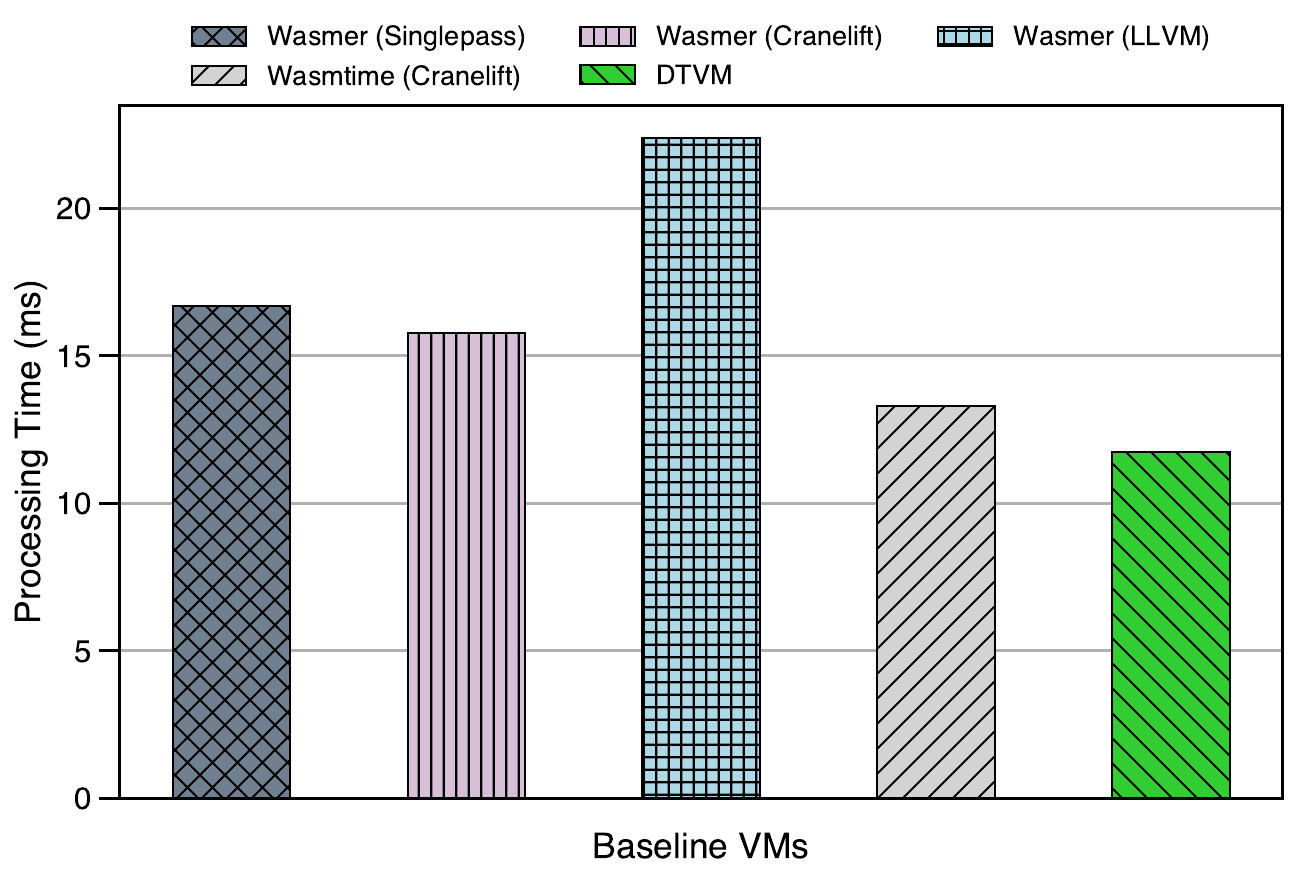}}
    \hspace{0.05in} 
    \subfigure[Integer Overflow]{\label{fig: exp1.2.2b}
    \includegraphics[width=0.4\columnwidth]{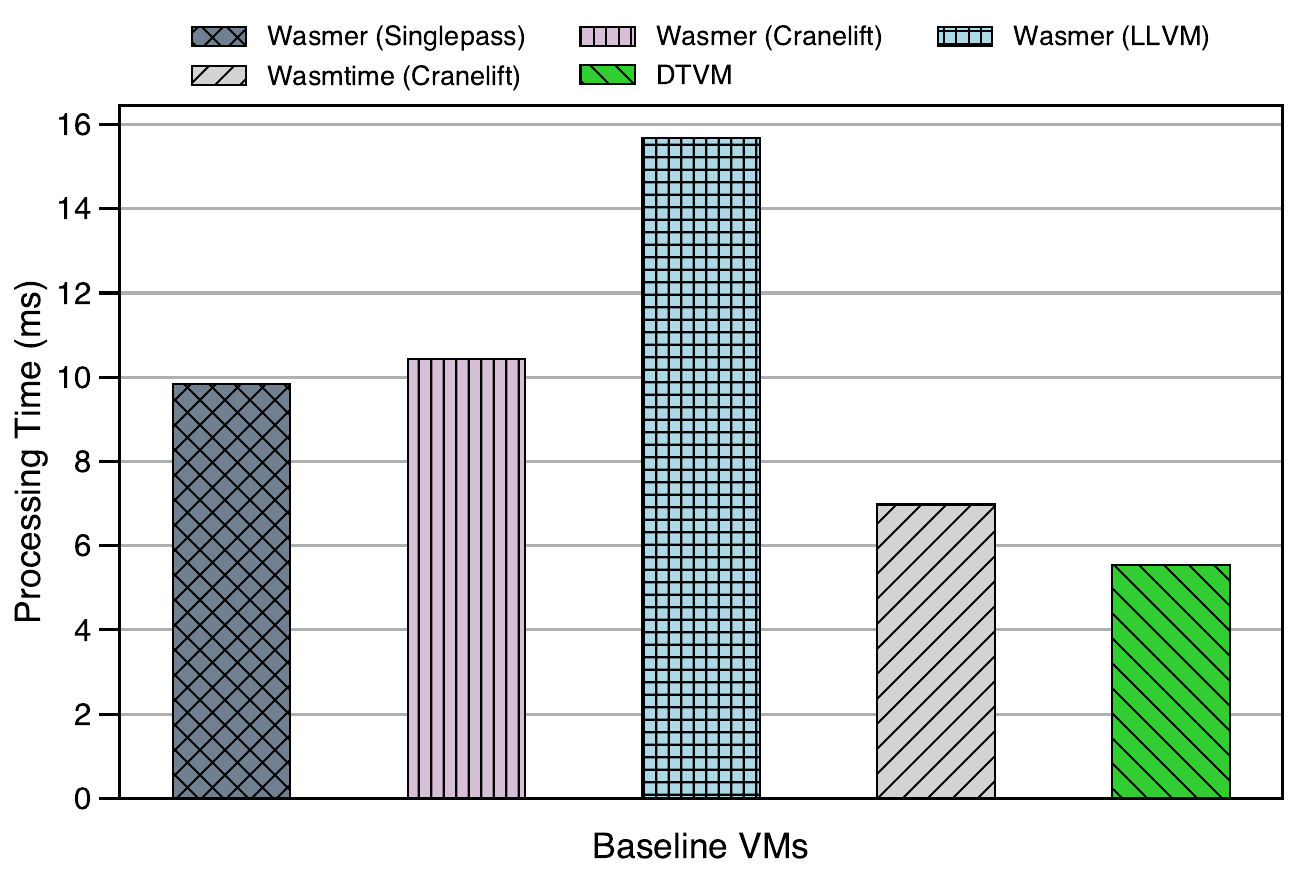}}
    \caption{Comparison with general-purpose Wasm Vms} \label{fig: exp1.2.2}
    \vspace{16pt}
\end{figure}

\textbf{Fibonacci: } 
As depicted in Fig.\ref{fig: exp1.2.2a}, for the Fibonacci sequence with a length of 30, \tool demonstrates the fastest processing time (11.73 ms), reducing the processing time by 11.8\%, 25.6\%, 29.6\%, and 40.5\% compared to Wasmtime (Cranelift), Wasmer (Cranelift), Wasmer (Singlepass), and Wasmer (LLVM), respectively. 


\textbf{Integer Overflow: } 
The processing time results for the Integer Overflow scenario are displayed in Fig.~\ref{fig: exp1.2.2b}. As can be seen, \tool has a processing time of only 5.5ms, delivering significant performance improvements compared to other baseline VMs. More specifically, it reduces processing time cost by nearly 20.6\% compared to Wasmtime (Cranelift), and up to 64.6\% compared to Wasmer variants.

\textbf{PolyBench: } We evaluate the runtime performance on the PolyBench benchmark, recording the processing time for each test case. Regarding the PolyBench consists of 30 scenarios, we present partial in Fig.~\ref{fig: exp1.2.2c} and the complete results are attached in Appendix~\ref{appx: polybench}. As depicted, \tool achieves notable reductions in processing time compared to other VMs, with 1.41$\times$ $\sim$ 1.63$\times$ improvements than Wasmtime (Cranelift). 

For a more detailed comparison, we provide an overview of processing time comparison, presented in Table~\ref{tab: Processing Time Comparison Overview on PolyBench Benchmark}. It is evident that \tool surpasses Wasmtime in over 66\% of PolyBench benchmark cases, with an average speedup of 1.20$\times$. Through further analysis of the remaining Cons cases of \tool, we observe that these scenarios primarily involve simple computations, though \tool exhibits clear advantages in compute-intensive cases, the further optimizations towards these simple cases will be included in the future work. When compared to all three Wasmer variants, \tool consistently outperforms across 100\% of the PolyBench cases, with an impressive average speedup ranging from 11.03$\times$ to 11.72$\times$. These results highlight \tool's superior efficiency in handling diverse compute-intensive workloads.

\begin{figure}[h]
    \setlength{\abovecaptionskip}{0cm}
    \setlength{\belowcaptionskip}{-1.5\baselineskip}
    \centering
    \includegraphics[width=0.7\columnwidth]{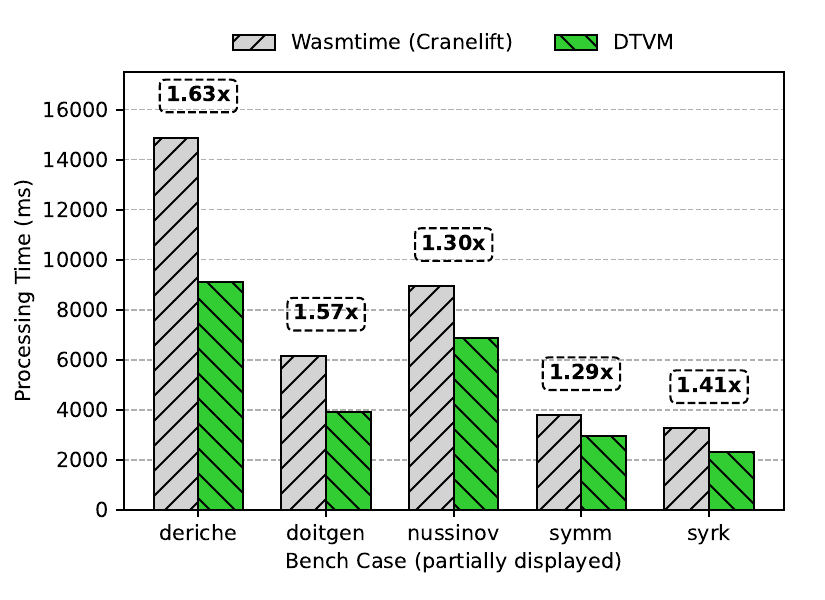}
    \caption{Results of PolyBench benchmark}\label{fig: exp1.2.2c}
\end{figure}

\begin{table}[h]
\centering
\small
\renewcommand{\arraystretch}{1.3}
\caption{Overview of PolyBench processing time comparison} \label{tab: Processing Time Comparison Overview on PolyBench Benchmark}
\vspace{5pt}
\begin{tabularx}{\textwidth}{p{5cm}XX}
    \toprule
   \makecell[l]{\textbf{Comparison Pair}}
 & \makecell[c]{\textbf{Pros Cases/Cons Cases}} 
 & \makecell[c]{\textbf{Average Speedup (Pros/Cons)}} \\
     \midrule
     \makecell[l]{\textbf{\tool vs. Wasmtime (Cranelift)}} & \makecell[c]{\textcolor{ForestGreen}{20} / \textcolor{Maroon}{10}} & \makecell[c]{1.20$\times$ / 0.76$\times$} \\
     \makecell[l]{\textbf{\tool vs. Wasmer (Singlepass)}} & \makecell[c]{\textcolor{ForestGreen}{30} / \textcolor{Maroon}{0}} & \makecell[c]{11.03$\times$ / ---}\\
     \makecell[l]{\textbf{\tool vs. Wasmer (Cranelift)}} & \makecell[c]{\textcolor{ForestGreen}{30} / \textcolor{Maroon}{0}} & \makecell[c]{12.14$\times$ / ---}\\
     \makecell[l]{\textbf{\tool vs. Wasmer (LLVM)}} & \makecell[c]{\textcolor{ForestGreen}{30} / \textcolor{Maroon}{0}} & \makecell[c]{11.72$\times$ / ---}\\
    \bottomrule
\end{tabularx}
\end{table}

\begin{figure}[h]
    \setlength{\abovecaptionskip}{0cm}
    \setlength{\belowcaptionskip}{-1.5\baselineskip}
    \centering
    \includegraphics[width=0.7\columnwidth]{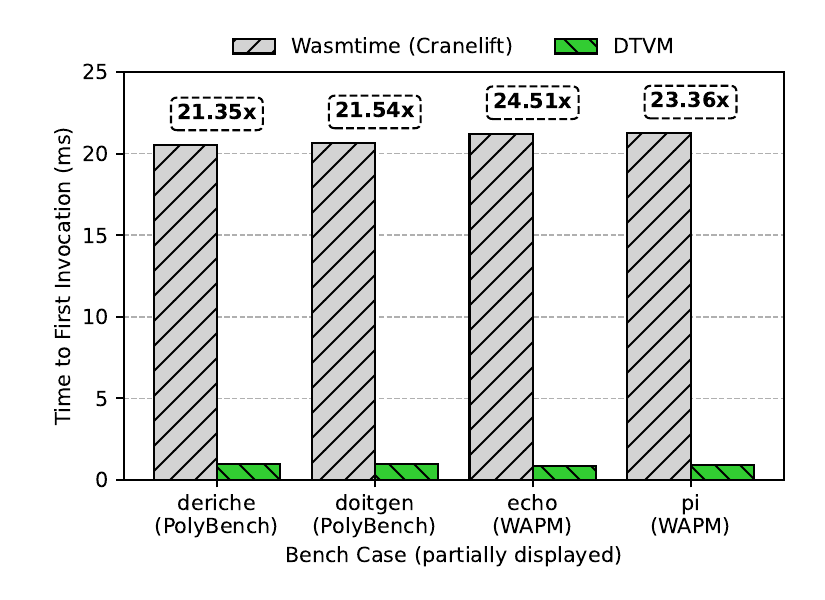}
    \caption{Comparsion of latency to first invocation}\label{fig: exp1.2.2d}
\end{figure}

\begin{table}[h]
\centering
\small
\renewcommand{\arraystretch}{1.3}
\caption{Overview of latency to first invocation comparison} \label{tab: Overview of latency to first invocation comparison}
\vspace{5pt}
\begin{tabularx}{\textwidth}{p{5cm}XX}
 \multicolumn{3}{c}{\textbf{PolyBench Benchmark}} \\
    \toprule
   \makecell[l]{\textbf{Comparison Pair}}
 & \makecell[c]{\textbf{Pros Cases/Cons Cases}} 
 & \makecell[c]{\textbf{Average Speedup (Pros/Cons)}} \\
     \midrule
     \makecell[l]{\textbf{\tool vs. Wasmtime (Cranelift)}} & \makecell[c]{\textcolor{ForestGreen}{30} / \textcolor{Maroon}{0}} & \makecell[c]{22.21$\times$ / ---}\\
     \makecell[l]{\textbf{\tool vs. Wasmer (Singlepass)}} & \makecell[c]{\textcolor{ForestGreen}{30} / \textcolor{Maroon}{0}} & \makecell[c]{4.05$\times$  / ---}\\
     \makecell[l]{\textbf{\tool vs. Wasmer (Cranelift)}} & \makecell[c]{\textcolor{ForestGreen}{30} / \textcolor{Maroon}{0}} & \makecell[c]{68.01$\times$  / ---}\\
     \makecell[l]{\textbf{\tool vs. Wasmer (LLVM)}} & \makecell[c]{\textcolor{ForestGreen}{30} / \textcolor{Maroon}{0}} & \makecell[c]{234.13$\times$  / ---}\\
     \midrule
     \\
     
     \multicolumn{3}{c}{\textbf{WAPM Benchmark}} \\
     \midrule
     \makecell[l]{\textbf{Comparison Pair}}
     & \makecell[c]{\textbf{Pros Cases/Cons Cases}} 
     & \makecell[c]{\textbf{Average Speedup (Pros/Cons)}} \\
     \midrule
     \makecell[l]{\textbf{\tool vs. Wasmtime (Cranelift)}} & \makecell[c]{\textcolor{ForestGreen}{30} / \textcolor{Maroon}{0}} & \makecell[c]{18.87$\times$ / ---}\\
     \makecell[l]{\textbf{\tool vs. Wasmer (Singlepass)}} & \makecell[c]{\textcolor{ForestGreen}{29} / \textcolor{Maroon}{1}} & \makecell[c]{4.35$\times$ / 0.49$\times$}\\
     \makecell[l]{\textbf{\tool vs. Wasmer (Cranelift)}} & \makecell[c]{\textcolor{ForestGreen}{30} / \textcolor{Maroon}{0}} & \makecell[c]{18.66$\times$ / ---}\\
     \makecell[l]{\textbf{\tool vs. Wasmer (LLVM)}} & \makecell[c]{\textcolor{ForestGreen}{30} / \textcolor{Maroon}{0}} & \makecell[c]{197.99$\times$ / ---}\\
    \bottomrule
\end{tabularx}
\end{table}

\subsubsection{Evaluation of Trampoline Hot-Switch Execution} We analyze the effectiveness of \tool's proposed hot-switch execution mechanism (with lazy compilation option enabled). The latency to first invocation of various VMs are compared to showcase \tool's advantage in on-demand compilation and rapid startup. We conduct experiments using benchmark suits from Workload C, the corresponding results are shown in Fig.~\ref{fig: exp1.2.2d}.

In cases from PolyBench, \tool's latency to first invocation (0.95ms $\sim$ 0.96ms) achieves an order of magnitude faster than Wasmtime (Cranelift) (20.5ms $\sim$ 20.6ms), leading to a maximum speedup of 21.54$\times$. Similarly, in WAPM scenarios, \tool exhibits a consistent lower latency, achieving speedups of 23.36$\times$ to 24.51$\times$.

In addition, we provide a statistical overview that analyzes \tool's latency to first invocation compared to other VMs across two benchmark suites. As shown in Table~\ref{tab: Overview of latency to first invocation comparison}, \tool shows latency advantages across all cases from PolyBench cases, when comparing to other baselines, with speedups ranging from 4.05$\times$ to 234.13$\times$. For WAPM cases, though \tool incurs only one slower case than Wasmer (Singlepass), it consistently demonstrate notable advantages in all other comparisons, achieving speedup ratios from 4.15$\times$ to 197.99$\times$. The results further demonstrate the enhanced flexibility and higher efficiency of on-demand JIT compilation enabled by the trampoline switchable execution.

\section{Conclusion}
In this paper, we present the DTVM Stack, a next-generation smart contract execution framework designed to address critical challenges in blockchain performance, determinism, and developer accessibility. By integrating a function-level lazy JIT compilation engine with deterministic execution guarantees, the DTVM Stack achieves significant performance improvements up to 58$\times$ faster than traditional interpreters and outperforming state-of-the-art Wasm VMs in over 66\% of benchmark cases. The framework bridges the gap between high-performance execution and EVM ABI compatibility, enabling seamless cross-language interoperability (C/C++, Rust, Java, Go, and AssemblyScript) while preserving harmony with Ethereum ecosystem. Furthermore, the integration of TEE-native security mechanisms reduces trusted computing base (TCB) size by 52\% compared to Wasmtime, which enhances security without sacrificing efficiency. The AI-driven development and auditing framework streamlines the contract development lifecycle, leveraging LLMs and multi-agent workflows to automate code generation, vulnerability detection, and deployment.


DTVM's unified JIT engine architecture establishes a foundational framework for  variants of ISA deterministic execution (e.g. Wasm, EVM, RISC-V), through modular frontend adaptation and shared optimization backend. Future advancements will focus on three synergistic directions: 1) Enhancing the dMIR intermediate layer with native support for cryptographic primitives and hardware-accelerated instructions, 2) Extending JIT backend compatibility to emerging CPU architectures while preserving execution determinism, and 3) Developing zero-knowledge-optimized VM variants that integrate zk-SNARK-friendly features. These extensions will enable DTVM to bridge performance-critical smart contract execution with verifiable computation paradigms, particularly for AI/ML-enhanced blockchain ecosystems, while maintaining backward compatibility through progressive WebAssembly specification adoption.

Collectively, these contributions advance the scalability and usability of blockchain infrastructure, offering a robust foundation for decentralized applications requiring deterministic execution, high throughput, and cross-platform interoperability. By open-sourcing the DTVM Stack and sharing optimizations with the EVM community, we aim to foster broader innovation in Web3 ecosystems.
\section*{Acknowledgment}
The successful development of DTVM Stack has been facilitated through collaborative contributions spanning multiple disciplines. We extend appreciation to Zhuoqun Bian, Chenguang Zhang, and Likai Wang for steering product strategy and architectural alignment with industry demands. Special recognition is given to Xuming Lu, Shan Yu, Rong Cao, Yunhan Hu, Jiangyi Chen, Yuyao Zhang, Linpeng Lu, Zhichuang Liang, Zijing Li, Yao Wang, Jie Mei, Zucheng Huang, and Li Lin for their engineering contributions, particularly in infrastructure optimization. 
Gratitude is expressed to Jiashui Wang and Lei Wang for their expert guidance in security penetration testing and secure implementation practices. Special acknowledgment is extended to Wenting Chang and Jin Peng for advancing standardization initiatives critical to DTVM's deterministic execution specifications. Recognition is also given to Lihua Zhang and his team for their dedication to engineering quality assurance.
Recognition is extended to Yajing Zhai, Wenting Ma, Yangtao Qiu, and Yingying Li for their strategic marketing and brand development initiatives. We would also like to extend our gratitude to Sikang Bian and Xu Wang for the contributions and support in the project's open-source efforts. Furthermore, gratitude is expressed to Prof. Aoying Zhou, Prof. Cheqing Jin, and Prof. Zhao Zhang, whose scholarly insights significantly enhanced theoretical rigor through academic collaborations.

\clearpage
\bibliographystyle{unsrt}  
\bibliography{ref}

\clearpage
\appendix
\section*{Appendix}

\section{PolyBench Results} \label{appx: polybench}
PolyBench is a benchmark suite of 30 numerical computations with
static control flow, extracted from operations in various application
domains, detailed in Table~\ref{tab: PolyBench Benchmark Descriptions}.

We measure the processing time and latency to first invocation across all cases of PolyBenchC benchmark (compiled in Wasm modules) under identical configurations, and the results are summarized in Table~\ref{tab: polybench}.

\begin{table}[ht]
\centering
\caption{PolyBench Benchmark Descriptions}
\vspace{5pt}
\label{tab: PolyBench Benchmark Descriptions}
\begin{tabularx}{\textwidth}{>{\bfseries}p{5cm}
>{\raggedright\arraybackslash}X}
\hline
\textbf{Benchmark Cases} & \textbf{Description} \\
\hline
2mm         & 2 Matrix Multiplications (alpha * A * B * C + beta * D) \\
3mm         & 3 Matrix Multiplications ((A*B)*(C*D)) \\
adi         & Alternating Direction Implicit solver \\
atax        & Matrix Transpose and Vector Multiplication \\
bicg        & BiCG Sub Kernel of BiCGStab Linear Solver \\
cholesky    & Cholesky Decomposition \\
correlation & Correlation Computation \\
covariance  & Covariance Computation \\
deriche     & Edge detection filter \\
doitgen     & Multi-resolution analysis kernel (MADNESS) \\
durbin      & Toeplitz system solver \\
fdtd-2d     & 2-D Finite Different Time Domain Kernel \\
floyd       & floyd-warshall algorithm \\
gemm        & Matrix-multiply C=alpha.A.B+beta.C \\
gemver      & Vector Multiplication and Matrix Addition \\
gesummv     & Scalar, Vector and Matrix Multiplication \\
gramschmidt & Gram-Schmidt decomposition \\
head-3d     & Heat equation over 3D data domain \\
jacobi-1d   & 1-D Jacobi stencil computation \\
jacobi-2d   & 2-D Jacobi stencil computation \\
lu          & LU decomposition \\
ludcmp      & LU decomposition followed by Forward Substitution \\
mvt         & Matrix Vector Product and Transpose \\
nussinov    & Dynamic programming algorithm for sequence alignment \\
seidel      & 2-D Seidel stencil computation \\
symm        & Symmetric matrix-multiply \\
syr2k       & Symmetric rank-2k update \\
syrk        & Symmetric rank-k update \\
trisolv     & Triangular solver \\
trmm        & Triangular matrix-multiply \\
\hline
\end{tabularx}
\end{table}

\begin{table}[h]
\centering
\small
\caption{Processing time of PolyBench (ms)}
\label{tab: polybench}
\vspace{5pt}
\begin{tabularx}{\textwidth}{
>{\bfseries}X
>{\centering\arraybackslash}X
>{\centering\arraybackslash}X
>{\centering\arraybackslash}X
>{\centering\arraybackslash}X
>{\centering\arraybackslash}X
>{\centering\arraybackslash}p{3.0cm}}
\toprule
\textbf{\makecell[c]{Bench Cases}} & 
{\textbf{\makecell[c]{\tool}}} & 
{\textbf{\makecell[c]{Wasmtime}}} & 
\textbf{\makecell[c]{Wasmer\\(Cranelift)}} & 
\textbf{\makecell[c]{Wasmer\\(LLVM)}} & 
\textbf{\makecell[c]{Wasmer\\(SinglePass)}} &
\textbf{\makecell[c]{\tool \\ Speedup Ratio (times)}}\\
\midrule

\multicolumn{7}{c}{\textbf{Processing Time}} \\

\midrule
2mm & 3283.6 & 3831.1 & 43503.2 & 43883.2 & 53043.6 & \textcolor{ForestGreen}{1.2}/\textcolor{ForestGreen}{13.2}/\textcolor{ForestGreen}{13.4}/\textcolor{ForestGreen}{16.2}\\

3mm & 4356.0 & 4888.1 & 41429.1 & 41364.7 & 57017.8 & \textcolor{ForestGreen}{1.1}/\textcolor{ForestGreen}{9.5}/\textcolor{ForestGreen}{9.5}/\textcolor{ForestGreen}{13.1}\\

adi & 11074.0 & 11689.5 & 53307.3 & 53055.3 & 63770.5 & \textcolor{ForestGreen}{1.1}/\textcolor{ForestGreen}{4.8}/\textcolor{ForestGreen}{4.8}/\textcolor{ForestGreen}{5.8}\\

atax & 87.5 & 63.2 & 153.7 & 339.8 & 196.6 & \textcolor{Maroon}{0.7}/\textcolor{ForestGreen}{1.8}/\textcolor{ForestGreen}{3.9}/\textcolor{ForestGreen}{2.2}\\

bicg & 88.1 & 66.1 & 230.3 & 437.0 & 272.9 & \textcolor{Maroon}{0.7}/\textcolor{ForestGreen}{2.6}/\textcolor{ForestGreen}{5.0}/\textcolor{ForestGreen}{3.1}\\

cholesky & 31295.1 & 31504.0 & 115307.6 & 115344.5 & 136898.5 & \textcolor{ForestGreen}{1.0}/\textcolor{ForestGreen}{3.7}/\textcolor{ForestGreen}{3.7}/\textcolor{ForestGreen}{4.4}\\

correlation & 8874.7 & 9211.4 & 69252.0 & 69184.2 & 71874.4 & \textcolor{ForestGreen}{1.0}/\textcolor{ForestGreen}{7.8}/\textcolor{ForestGreen}{7.8}/\textcolor{ForestGreen}{8.1}\\

covariance & 9072.9 & 10174.1 & 70404.7 & 68993.9 & 70880.9 & \textcolor{ForestGreen}{1.1}/\textcolor{ForestGreen}{7.8}/\textcolor{ForestGreen}{7.6}/\textcolor{ForestGreen}{7.8}\\

deriche & 9115.0 & 14877.2 & 382532.6 & 378533.2 & 336237.2 & \textcolor{ForestGreen}{1.6}/\textcolor{ForestGreen}{42.0}/\textcolor{ForestGreen}{41.5}/\textcolor{ForestGreen}{36.9}\\

doitgen & 3933.6 & 6158.0 & 145150.0 & 146429.3 & 106225.9 & \textcolor{ForestGreen}{1.6}/\textcolor{ForestGreen}{36.9}/\textcolor{ForestGreen}{37.2}/\textcolor{ForestGreen}{27.0}\\

durbin & 44.9 & 38.9 & 154.9 & 373.3 & 133.4 & \textcolor{Maroon}{0.9}/\textcolor{ForestGreen}{3.5}/\textcolor{ForestGreen}{8.3}/\textcolor{ForestGreen}{3.0}\\

fdtd-2d & 6473.0 & 7877.1 & 157737.1 & 156527.0 & 164308.5 & \textcolor{ForestGreen}{1.2}/\textcolor{ForestGreen}{24.4}/\textcolor{ForestGreen}{24.2}/\textcolor{ForestGreen}{25.4}\\

floyd & 26208.4 & 30132.1 & 352050.5 & 352885.4 & 426687.5 & \textcolor{ForestGreen}{1.1}/\textcolor{ForestGreen}{13.4}/\textcolor{ForestGreen}{13.5}/\textcolor{ForestGreen}{16.3}\\

gemm & 2661.9 & 2984.6 & 48471.9 & 45035.0 & 51762.5 & \textcolor{ForestGreen}{1.1}/\textcolor{ForestGreen}{18.2}/\textcolor{ForestGreen}{16.9}/\textcolor{ForestGreen}{19.4}\\

gemver & 103.8 & 76.1 & 168.1 & 362.1 & 228.2 & \textcolor{Maroon}{0.7}/\textcolor{ForestGreen}{1.6}/\textcolor{ForestGreen}{3.5}/\textcolor{ForestGreen}{2.2}\\

gesummv & 80.5 & 50.9 & 112.5 & 306.4 & 126.1 & \textcolor{Maroon}{0.6}/\textcolor{ForestGreen}{1.4}/\textcolor{ForestGreen}{3.8}/\textcolor{ForestGreen}{1.6}\\

gramschmidt & 11168.4 & 13310.2 & 123374.2 & 122096.3 & 125087.2 & \textcolor{ForestGreen}{1.2}/\textcolor{ForestGreen}{11.0}/\textcolor{ForestGreen}{10.9}/\textcolor{ForestGreen}{11.2}\\

head-3d & 7026.7 & 6270.8 & 78010.4 & 77935.5 & 89053.8 & \textcolor{Maroon}{0.9}/\textcolor{ForestGreen}{11.1}/\textcolor{ForestGreen}{11.1}/\textcolor{ForestGreen}{12.7}\\

jacobi-2d & 4792.0 & 4964.5 & 75313.1 & 74517.9 & 81685.0 & \textcolor{ForestGreen}{1.0}/\textcolor{ForestGreen}{15.7}/\textcolor{ForestGreen}{15.6}/\textcolor{ForestGreen}{17.0}\\

jacobi-1d & 53.0 & 36.5 & 127.8 & 321.2 & 113.9 & \textcolor{Maroon}{0.7}/\textcolor{ForestGreen}{2.4}/\textcolor{ForestGreen}{6.1}/\textcolor{ForestGreen}{2.1}\\

lu & 36298.0 & 38748.8 & 204999.7 & 204062.8 & 234819.0 & \textcolor{ForestGreen}{1.1}/\textcolor{ForestGreen}{5.6}/\textcolor{ForestGreen}{5.6}/\textcolor{ForestGreen}{6.5}\\

ludcmp & 32175.7 & 31975.8 & 32541.2 & 32506.1 & 56426.4 & \textcolor{Maroon}{1.0}/\textcolor{ForestGreen}{1.0}/\textcolor{ForestGreen}{1.0}/\textcolor{ForestGreen}{1.8}\\

mvt & 103.9 & 75.6 & 246.6 & 451.7 & 298.4 & \textcolor{Maroon}{0.7}/\textcolor{ForestGreen}{2.4}/\textcolor{ForestGreen}{4.3}/\textcolor{ForestGreen}{2.9}\\

nussinov & 6866.0 & 8939.5 & 138454.1 & 138719.1 & 141543.7 & \textcolor{ForestGreen}{1.3}/\textcolor{ForestGreen}{20.2}/\textcolor{ForestGreen}{20.2}/\textcolor{ForestGreen}{20.6}\\

seidel & 25773.3 & 28268.1 & 194523.5 & 194801.1 & 137644.3 & \textcolor{ForestGreen}{1.1}/\textcolor{ForestGreen}{7.5}/\textcolor{ForestGreen}{7.6}/\textcolor{ForestGreen}{5.3}\\

symm & 2942.1 & 3780.7 & 53122.2 & 53885.1 & 57868.0 & \textcolor{ForestGreen}{1.3}/\textcolor{ForestGreen}{18.1}/\textcolor{ForestGreen}{18.3}/\textcolor{ForestGreen}{19.7}\\

syr2k & 5653.1 & 6585.4 & 51200.4 & 65982.3 & 68709.0 & \textcolor{ForestGreen}{1.2}/\textcolor{ForestGreen}{9.1}/\textcolor{ForestGreen}{11.7}/\textcolor{ForestGreen}{12.2}\\

syrk & 2319.8 & 3260.9 & 37506.9 & 62935.5 & 65630.0 & \textcolor{ForestGreen}{1.4}/\textcolor{ForestGreen}{16.2}/\textcolor{ForestGreen}{27.1}/\textcolor{ForestGreen}{28.3}\\

trisolv & 82.0 & 47.1 & 142.7 & 335.5 & 152.3 & \textcolor{Maroon}{0.6}/\textcolor{ForestGreen}{1.7}/\textcolor{ForestGreen}{4.1}/\textcolor{ForestGreen}{1.9}\\

trmm & 3287.4 & 4037.9 & 54060.7 & 53542.7 & 56432.5 & \textcolor{ForestGreen}{1.2}/\textcolor{ForestGreen}{16.4}/\textcolor{ForestGreen}{16.3}/\textcolor{ForestGreen}{17.2}\\

\midrule
\multicolumn{7}{c}{\textbf{Latency to First Invocation}} \\

\midrule

2mm & 0.9 & 20.4 & 19.6 & 222.8 & 3.4 & \textcolor{ForestGreen}{21.8}/\textcolor{ForestGreen}{21.0}/\textcolor{ForestGreen}{238.3}/\textcolor{ForestGreen}{3.7}\\

3mm & 1.0 & 20.6 & 19.8 & 215.5 & 3.4 & \textcolor{ForestGreen}{20.7}/\textcolor{ForestGreen}{20.0}/\textcolor{ForestGreen}{217.0}/\textcolor{ForestGreen}{3.4}\\

adi & 1.0 & 19.2 & 22.9 & 225.5 & 3.3 & \textcolor{ForestGreen}{20.1}/\textcolor{ForestGreen}{24.0}/\textcolor{ForestGreen}{235.9}/\textcolor{ForestGreen}{3.4}\\

atax & 0.9 & 20.2 & 19.7 & 208.5 & 3.6 & \textcolor{ForestGreen}{23.5}/\textcolor{ForestGreen}{22.9}/\textcolor{ForestGreen}{242.7}/\textcolor{ForestGreen}{4.1}\\

bicg & 0.9 & 22.4 & 19.3 & 221.7 & 3.8 & \textcolor{ForestGreen}{24.9}/\textcolor{ForestGreen}{21.5}/\textcolor{ForestGreen}{246.6}/\textcolor{ForestGreen}{4.2}\\

cholesky & 1.0 & 19.8 & 18.5 & 215.7 & 3.7 & \textcolor{ForestGreen}{20.7}/\textcolor{ForestGreen}{19.4}/\textcolor{ForestGreen}{225.9}/\textcolor{ForestGreen}{3.9}\\

correlation & 1.0 & 20.7 & 19.7 & 227.2 & 4.2 & \textcolor{ForestGreen}{21.3}/\textcolor{ForestGreen}{20.2}/\textcolor{ForestGreen}{233.5}/\textcolor{ForestGreen}{4.3}\\

covariance & 0.9 & 20.9 & 19.4 & 208.1 & 4.0 & \textcolor{ForestGreen}{22.3}/\textcolor{ForestGreen}{20.7}/\textcolor{ForestGreen}{221.9}/\textcolor{ForestGreen}{4.3}\\

deriche & 1.0 & 20.5 & 20.8 & 209.9 & 4.0 & \textcolor{ForestGreen}{21.4}/\textcolor{ForestGreen}{21.6}/\textcolor{ForestGreen}{218.2}/\textcolor{ForestGreen}{4.2}\\

doitgen & 1.0 & 20.7 & 24.3 & 234.0 & 4.4 & \textcolor{ForestGreen}{21.5}/\textcolor{ForestGreen}{25.3}/\textcolor{ForestGreen}{244.0}/\textcolor{ForestGreen}{4.6}\\

durbin & 0.9 & 20.3 & 19.8 & 236.3 & 5.0 & \textcolor{ForestGreen}{22.3}/\textcolor{ForestGreen}{21.8}/\textcolor{ForestGreen}{259.7}/\textcolor{ForestGreen}{5.5}\\

fdtd-2d & 0.9 & 19.8 & 20.3 & 214.8 & 4.2 & \textcolor{ForestGreen}{22.2}/\textcolor{ForestGreen}{22.8}/\textcolor{ForestGreen}{241.4}/\textcolor{ForestGreen}{4.8}\\

floyd & 1.0 & 20.1 & 19.5 & 302.5 & 3.6 & \textcolor{ForestGreen}{20.0}/\textcolor{ForestGreen}{19.4}/\textcolor{ForestGreen}{300.7}/\textcolor{ForestGreen}{3.6}\\

gemm & 0.9 & 20.9 & 19.7 & 211.3 & 3.3 & \textcolor{ForestGreen}{22.6}/\textcolor{ForestGreen}{21.4}/\textcolor{ForestGreen}{228.9}/\textcolor{ForestGreen}{3.6}\\

gemver & 0.9 & 20.6 & 19.4 & 210.2 & 3.8 & \textcolor{ForestGreen}{24.0}/\textcolor{ForestGreen}{22.6}/\textcolor{ForestGreen}{244.7}/\textcolor{ForestGreen}{4.4}\\

gesummv & 0.9 & 20.1 & 20.1 & 209.6 & 3.3 & \textcolor{ForestGreen}{21.4}/\textcolor{ForestGreen}{21.4}/\textcolor{ForestGreen}{223.2}/\textcolor{ForestGreen}{3.6}\\

gramschmidt & 1.0 & 20.1 & 22.2 & 211.2 & 3.2 & \textcolor{ForestGreen}{20.5}/\textcolor{ForestGreen}{22.6}/\textcolor{ForestGreen}{215.3}/\textcolor{ForestGreen}{3.3}\\

head-3d & 0.9 & 23.4 & 19.6 & 228.0 & 3.4 & \textcolor{ForestGreen}{25.2}/\textcolor{ForestGreen}{21.0}/\textcolor{ForestGreen}{245.5}/\textcolor{ForestGreen}{3.7}\\

jacobi-2d & 1.0 & 20.9 & 19.6 & 220.1 & 3.5 & \textcolor{ForestGreen}{21.9}/\textcolor{ForestGreen}{20.5}/\textcolor{ForestGreen}{230.5}/\textcolor{ForestGreen}{3.6}\\

jacobi-1d & 1.0 & 20.3 & 19.9 & 215.4 & 3.4 & \textcolor{ForestGreen}{20.6}/\textcolor{ForestGreen}{20.2}/\textcolor{ForestGreen}{218.5}/\textcolor{ForestGreen}{3.5}\\

lu & 0.9 & 22.9 & 18.7 & 205.0 & 3.1 & \textcolor{ForestGreen}{24.8}/\textcolor{ForestGreen}{20.3}/\textcolor{ForestGreen}{222.3}/\textcolor{ForestGreen}{3.3}\\

ludcmp & 0.9 & 21.4 & 19.4 & 202.0 & 3.4 & \textcolor{ForestGreen}{22.7}/\textcolor{ForestGreen}{20.6}/\textcolor{ForestGreen}{214.0}/\textcolor{ForestGreen}{3.6}\\

mvt & 0.9 & 20.8 & 19.5 & 222.0 & 3.5 & \textcolor{ForestGreen}{22.6}/\textcolor{ForestGreen}{21.2}/\textcolor{ForestGreen}{241.0}/\textcolor{ForestGreen}{3.8}\\

nussinov & 0.9 & 20.1 & 19.8 & 210.9 & 3.3 & \textcolor{ForestGreen}{22.6}/\textcolor{ForestGreen}{22.2}/\textcolor{ForestGreen}{237.2}/\textcolor{ForestGreen}{3.7}\\

seidel & 0.9 & 20.4 & 19.9 & 211.6 & 3.5 & \textcolor{ForestGreen}{21.5}/\textcolor{ForestGreen}{20.9}/\textcolor{ForestGreen}{222.9}/\textcolor{ForestGreen}{3.7}\\

symm & 1.0 & 22.8 & 20.3 & 212.6 & 5.4 & \textcolor{ForestGreen}{23.8}/\textcolor{ForestGreen}{21.2}/\textcolor{ForestGreen}{221.9}/\textcolor{ForestGreen}{5.6}\\

syr2k & 0.9 & 20.2 & 29.3 & 213.1 & 4.2 & \textcolor{ForestGreen}{22.1}/\textcolor{ForestGreen}{32.1}/\textcolor{ForestGreen}{233.4}/\textcolor{ForestGreen}{4.6}\\

syrk & 0.9 & 20.8 & 22.1 & 213.4 & 4.2 & \textcolor{ForestGreen}{22.4}/\textcolor{ForestGreen}{23.8}/\textcolor{ForestGreen}{230.4}/\textcolor{ForestGreen}{4.5}\\

trisolv & 0.9 & 20.7 & 23.9 & 215.8 & 4.2 & \textcolor{ForestGreen}{23.0}/\textcolor{ForestGreen}{26.6}/\textcolor{ForestGreen}{239.8}/\textcolor{ForestGreen}{4.7}\\

trmm & 0.9 & 20.0 & 24.5 & 210.3 & 3.9 & \textcolor{ForestGreen}{21.8}/\textcolor{ForestGreen}{26.6}/\textcolor{ForestGreen}{228.8}/\textcolor{ForestGreen}{4.3}\\

\bottomrule
\end{tabularx}
\end{table}



\clearpage
\section{WAPM Results} \label{appx: wapm}

\begin{table}[h]
\centering
\small
\caption{Latency to first invocation of WAPM (ms)}
\label{tab: wapm_compilation_time}
\vspace{5pt}
\begin{tabularx}{\textwidth}{
>{\bfseries}X
>{\centering\arraybackslash}X
>{\centering\arraybackslash}X
>{\centering\arraybackslash}X
>{\centering\arraybackslash}X
>{\centering\arraybackslash}X
>{\centering\arraybackslash}p{3cm}}
\toprule
\textbf{\makecell[c]{Bench Cases}} & 
{\textbf{\makecell[c]{\tool}}} & 
{\textbf{\makecell[c]{Wasmtime}}} & 
\textbf{\makecell[c]{Wasmer\\(Cranelift)}} & 
\textbf{\makecell[c]{Wasmer\\(LLVM)}} & 
\textbf{\makecell[c]{Wasmer\\(SinglePass)}} &
\textbf{\makecell[c]{\tool \\ Speedup Ratio (times)}}\\
\midrule

amirali & 0.4 & 2.2 & 1.5 & 10.2 & 1.1 & \textcolor{ForestGreen}{6.1}/\textcolor{ForestGreen}{4.3}/\textcolor{ForestGreen}{28.5}/\textcolor{ForestGreen}{2.9}\\

base64cli & 5.3 & 120.5 & 115.0 & 1394.9 & 29.0 & \textcolor{ForestGreen}{22.7}/\textcolor{ForestGreen}{21.6}/\textcolor{ForestGreen}{262.4}/\textcolor{ForestGreen}{5.5}\\

chkfont & 1.4 & 28.7 & 39.1 & 284.6 & 5.0 & \textcolor{ForestGreen}{20.3}/\textcolor{ForestGreen}{27.7}/\textcolor{ForestGreen}{201.1}/\textcolor{ForestGreen}{3.5}\\

code-stock & 1.4 & 22.0 & 35.0 & 270.3 & 9.2 & \textcolor{ForestGreen}{15.9}/\textcolor{ForestGreen}{25.2}/\textcolor{ForestGreen}{194.6}/\textcolor{ForestGreen}{6.6}\\

cowsay & 4.6 & 111.1 & 106.5 & 1312.0 & 22.3 & \textcolor{ForestGreen}{24.2}/\textcolor{ForestGreen}{23.2}/\textcolor{ForestGreen}{286.3}/\textcolor{ForestGreen}{4.9}\\

dice & 0.8 & 11.6 & 10.5 & 109.3 & 2.6 & \textcolor{ForestGreen}{14.0}/\textcolor{ForestGreen}{12.6}/\textcolor{ForestGreen}{131.7}/\textcolor{ForestGreen}{3.1}\\

echo & 0.9 & 21.2 & 20.8 & 223.5 & 3.3 & \textcolor{ForestGreen}{24.5}/\textcolor{ForestGreen}{24.0}/\textcolor{ForestGreen}{258.1}/\textcolor{ForestGreen}{3.8}\\

erdtree & 2.1 & 40.7 & 34.4 & 321.9 & 10.7 & \textcolor{ForestGreen}{19.1}/\textcolor{ForestGreen}{16.1}/\textcolor{ForestGreen}{150.8}/\textcolor{ForestGreen}{5.0}\\

figlet & 1.6 & 30.6 & 10.3 & 100.4 & 2.4 & \textcolor{ForestGreen}{19.0}/\textcolor{ForestGreen}{6.4}/\textcolor{ForestGreen}{62.3}/\textcolor{ForestGreen}{1.5}\\

irb & 270.6 & 6061.1 & 6530.3 & 64291.0 & 828.3 & \textcolor{ForestGreen}{22.4}/\textcolor{ForestGreen}{24.1}/\textcolor{ForestGreen}{237.6}/\textcolor{ForestGreen}{3.1}\\

md5 & 0.9 & 21.5 & 21.1 & 234.4 & 3.4 & \textcolor{ForestGreen}{22.9}/\textcolor{ForestGreen}{22.5}/\textcolor{ForestGreen}{249.3}/\textcolor{ForestGreen}{3.7}\\

pi & 0.9 & 21.3 & 20.9 & 230.4 & 3.3 & \textcolor{ForestGreen}{23.4}/\textcolor{ForestGreen}{22.9}/\textcolor{ForestGreen}{253.2}/\textcolor{ForestGreen}{3.6}\\

pkg1 & 1.3 & 21.4 & 25.3 & 262.7 & 9.1 & \textcolor{ForestGreen}{16.3}/\textcolor{ForestGreen}{19.3}/\textcolor{ForestGreen}{200.3}/\textcolor{ForestGreen}{6.9}\\

qr2text & 1.7 & 34.6 & 32.4 & 364.4 & 7.3 & \textcolor{ForestGreen}{20.0}/\textcolor{ForestGreen}{18.8}/\textcolor{ForestGreen}{210.7}/\textcolor{ForestGreen}{4.2}\\

sam & 1.2 & 24.6 & 24.3 & 274.7 & 4.9 & \textcolor{ForestGreen}{21.2}/\textcolor{ForestGreen}{21.0}/\textcolor{ForestGreen}{236.8}/\textcolor{ForestGreen}{4.2}\\

suggest & 4.9 & 112.8 & 106.2 & 1262.1 & 26.3 & \textcolor{ForestGreen}{22.8}/\textcolor{ForestGreen}{21.5}/\textcolor{ForestGreen}{255.5}/\textcolor{ForestGreen}{5.3}\\

tpl & 14.7 & 208.5 & 192.3 & 2199.5 & 58.6 & \textcolor{ForestGreen}{14.2}/\textcolor{ForestGreen}{13.1}/\textcolor{ForestGreen}{150.0}/\textcolor{ForestGreen}{4.0}\\

uuid & 5.3 & 114.7 & 115.5 & 1378.6 & 28.2 & \textcolor{ForestGreen}{21.7}/\textcolor{ForestGreen}{21.9}/\textcolor{ForestGreen}{261.0}/\textcolor{ForestGreen}{5.3}\\

viu & 13.2 & 309.6 & 312.4 & 3911.3 & 71.3 & \textcolor{ForestGreen}{23.5}/\textcolor{ForestGreen}{23.7}/\textcolor{ForestGreen}{297.2}/\textcolor{ForestGreen}{5.4}\\

zuk & 282.9 & 911.5 & 912.1 & 9197.4 & 139.7 & \textcolor{ForestGreen}{3.2}/\textcolor{ForestGreen}{3.2}/\textcolor{ForestGreen}{32.5}/\textcolor{Maroon}{0.5}\\


\bottomrule
\end{tabularx}
\end{table}

\end{document}